\documentclass[fleqn,10pt]{wlscirep}
\usepackage[utf8]{inputenc}
\usepackage[T1]{fontenc}
\usepackage{algorithm}
\usepackage{algpseudocode}
\usepackage{amsfonts}
\usepackage{amsmath}
\usepackage{caption}
\usepackage[T1]{fontenc}
\usepackage{hyperref}
\usepackage[utf8]{inputenc}
\usepackage{mathrsfs}
\usepackage{moreverb}
\usepackage{multicol}
\usepackage{derivative}

\newcommand{\bsym}[1]{\boldsymbol{#1}}

\newcommand\BibTeX{{\rmfamily B\kern-.05em \textsc{i\kern-.025em b}\kern-.08em
T\kern-.1667em\lower.7ex\hbox{E}\kern-.125emX}}

\renewcommand{\vec}[1]{\boldsymbol{#1}}

\title{Cross-validation of meshless Navier-Stokes solvers in porous media flows}

\author[1,2,*]{Dawid Strzelczyk}
\author[2,3]{Miha Rot}
\author[2]{Gregor Kosec}
\author[1,2]{Maciej Matyka}

\affil[1]{Faculty of Physics and Astronomy, University of Wroc\l aw, pl. Maxa Borna 9, 50-204 Wroc\l aw, Poland}
\affil[2]{Parallel and Distributed Systems Laboratory, Jo\v{z}ef Stefan Institute, Jamova cesta 39, 1000 Ljubljana, Slovenia}
\affil[3]{Jožef Stefan International Postgraduate School, Jamova cesta 39, 1000 Ljubljana, Slovenia}

\affil[*]{dawid.strzelczyk@uwr.edu.pl}


\begin{abstract}
In this paper, two mesh-free CFD solvers for pore-scale fluid flow through porous media are considered, namely the Lattice Boltzmann Method with the two relaxation time collision term and the direct Navier-Stokes solver under the artificial compressibility limit. The porous media is built with a regular arrangement of spherical grains with variable radii, which allows control of the porosity. Both solvers use the same $h$-refined meshless spatial discretization to adequately capture the underlying geometry and the same Radial Basis Function (RBF) method to approximate the involved fields and partial differential operators. First, the results are compared with the data from the literature in terms of drag coefficient and permeability at different porosities achieving excellent agreement with the reported results. Next, the simulations are extended beyond the porosity range reported in the literature using proposed $h$-refined CFD solvers. The results are supported by convergence and timing analyses and discussions on meshless parameters such as stencil size and refinement settings.
\end{abstract}
\begin{document}

\flushbottom
\maketitle

\thispagestyle{empty}

\section{Introduction}
Porous media are ubiquitous and the fluid flow through them is of paramount importance for science, technology, and life. It concerns such fundamental issues as the health and performance of the human body, which is composed of $70 \%$ fluids, where the lungs and the arterial system can indeed be treated as porous systems~\cite{miguel2011lungs}, or solidification processes, where the transport of solutes in mushy (porous) regime governs the macrosegregation effect~\cite{whitesell2001influence}, and after all, study of radioactive waste seepage in radioactive waste disposal facilities~\cite{patel2018three}, to name just a few examples.
 
Research in the field of fluid dynamics (porous or free fluid) is not possible with analytical methods and therefore requires sophisticated experimental methods, complex differential calculus methods, or the use of numerical methods. Traditionally, problems in fluid dynamics have been approached numerically using mesh-based methods, where, despite the long history of research on mesh generators~\cite{singh2023review}, meshing remains a difficult problem that cannot be fully automated and therefore often requires significant human assistance. As a matter of fact, the meshing of irregularly shaped 3D geometries, especially when dealing with complex geometries such as porous media~\cite{klostermann2013meshing}, is one of the most complex and time-consuming steps in the overall mesh-based numerical solution~\cite{liu2005introduction}.

In response to the complexity of meshing, the development of numerical methods took two principal directions. In addition to the development of specific methods and algorithms for meshing~\cite{singh2023review}, a class of numerical methods based on the meshless principle has been developed~\cite{nguyen_meshless_2008}. The conceptual difference between mesh-based and meshless methods lies in the consideration of the relationships between the computational nodes. Meshless methods define the relationship between the nodes completely using only the internodal distances, thus freeing themselves from the constraints of using meshes. An important implication of this simplification is that meshless methods can work with scattered nodes. Although it is generally recognized that certain rules must be followed when generating such scattered nodes~\cite{Slak2019}, the positioning of the nodes is significantly less complex compared to meshing~\cite{zienkiewicz_finite_2005} and can be automated regardless of the dimensionality or shape of the domain under consideration~\cite{Slak2019, depolli_parallel_2022}. Moreover, the $h$-adaptivity in the meshless setup is almost free, since the nodes can be distributed with variable internodal distance without any special treatment~\cite{oanh_approach_2022, Slak2019, jancic_strong_2023}. In recent years, a specific meshless method has gained popularity, namely a radial basis function (RBF) generated finite differences (FD) ~\cite{tolstykh_using_2003}. The hallmark of the RBF-FD is an approximation based on the combination of polynomials that ensure consistency up to their order~\cite{Bayona2017} and RBF (typically polyharmonic splines) that help stabilize the method. Although the idea is not entirely new, recent theoretical and experimental research has taken the understanding of such a numerical approach to a new level~\cite{Bayona2017, slak_adaptive_2019, jancic_strong_2023, oanh_approach_2022, chu_rbf-fd_2023}.

In fluid dynamics, different meshless approaches have been successfully applied to the Navier-Stokes problem, namely methods based on direct solving of Navier-Stokes equation using different meshless approximations and pressure-velocity couplings~\cite{chu_rbf-fd_2023, chinchapatnam_compact_2009, zamolo2023accurate} and Meshless lattice Boltzmann methods~\cite{musavi2015meshless, Strzelczyk2024}. Each of these methods has different properties and is used differently for certain physical problems. It is important to understand and employ them in a proper way to utilize their strengths and avoid weaknesses in the given context, which in our case is the fluid flow through the irregular, complex pores of porosity-dependent porous media.

The aim of this work is to formulate and implement a direct meshless Navier-Stokes solver under the artificial compressibility limit and meshless Lattice Boltzmann Method with the two relaxation time collision. Both methods use the same $h$-refined meshless spatial discretization and the same RBF method to approximate the involved fields and partial differential operators~\cite{slak_medusa_2021} for a fair comparison. We test both methods on the classical problem of flow through idealized three-dimensional porous media represented as a periodic array of spheres and compare meshless results with the previously recognized benchmarks, i.e. smoothed particle hydrodynamics method based solution provided by Holmes, et al.~\cite{Holmes2011} and solution provided by Larson and Higdon~\cite{Larson1989}. We are particularly interested in extending the simulations to the limiting cases of the model, i.e. low porosity, high porosity, and touching limit of the singularity contact point. Under these conditions, we discuss the strengths and weaknesses of both formulations in terms of numerical accuracy and stability, convergence rate, complexity, and robustness using refined scattered nodes. Furthermore, we identify regions where both approaches have problems and discuss possible overlaps and gap-filling between the two. Finally, we extend the spectra of available benchmark data by expanding the range of porosity in the simulations.   

\section{Methods}\label{sec:methods}

We study the pore scale flow through a porous medium by modeling it with the incompressible Navier-Stokes system of equations
\begin{equation}\label{eq:NSeq}
    \begin{aligned}
        \nabla \cdot \vec{v} &= 0,\\
        \pdv{\vec{v}}{t} + (\vec{v} \cdot 
     \nabla) \vec{v} &= -\dfrac{1}{\rho} \nabla p + \nu \nabla^2\vec{v} + \vec{g}.
    \end{aligned}
\end{equation}
where $\bsym{v}(t,\bsym{x})$ is the velocity field, $p(t,\bsym{x})$ is the pressure field, and $\rho$, $\nu$ and $\bsym{g}$ are the density, the kinematic viscosity and the body force, respectively. To solve the equations, we take two approaches. The first is to solve the Boltzmann transport equation with the meshless Lattice Boltzmann Method (MLBM) described in Sec.~\ref{ssec:MLBM_method} and then, by the virtue of Chapman-Enskog expansion\cite{Krueger2016}, derive $\bsym{v}$ and $p$ fields from its solution. The other is to solve Eq.~\eqref{eq:NSeq} directly, using a meshless Navier-Stokes (MNS) solver described in Section.~\ref{ssec:MNS_method}. The pseudocodes for MNS and MLBM are provided in Appendix~\ref{app:algorithms_comparison}. Both methods rely on the meshless RBF-FD approximation framework; either to interpolate values between the Eulerian and the Lagrangian nodes in LBM or to approximate the spatial derivatives in MNS. The principles of the approximation method and the description of the node positioning algorithm used in the present study are provided in Sec.~\ref{ssec:meshless}.

\subsection{Meshless discretization and approximation}\label{ssec:meshless}
The discretization of the computational domain for the meshless approximation is achieved by positioning $N$ scattered \emph{computational nodes} $\vec{x}_i \in \mathbb{R}^D$, with $D$ denoting the spatial dimension. The nodes are placed with a variable density, defined with the inter-nodal distance function $h(\vec{x})$, utilizing a dedicated meshless dimension-independent variable density (DIVG) node positioning algorithm~\cite{slak2019generation} with an extension to account for the periodic nature of the domain. DIVG is an iterative algorithm based on an \emph{expansion queue} that results in discretization proceeding as an advancing front away from the initial nodes used to seed the queue. At each iteration, we attempt to expand the discretization with candidate nodes laying on a circle with radius $h(\vec{x}_\mathrm{d})$ around the de-queued node position $\vec{x}_\mathrm{d}$. The candidates that fall inside the domain and not too close to existing nodes, based on their local $h(\vec{x}_i)$, are then added to the discretization and to the queue for further expansion which continues until the queue is empty.

No further information, e.g. meshing, regarding the node cloud is required as the \emph{stencil} $S_i$, i.e. the domain for the local approximation in the $i$-th computational node, is constructed solely based on the inter-nodal distance. Although more complex strategies for stencil construction exist~\cite{davydov2023stencilSelection} we use the simplest and populate $S_i$ with indices of $N_L$ closest neighbors to the $i$-th computational node.

After the stencils for computational nodes are constructed we can use them to form a generalized finite difference approximation that numerically approximates the linear differential operator $\mathcal{L}$ in the central node 
\begin{equation}
	(\mathcal{L} u)_i \approx \sum_{j=1}^{N_L} w_{i, j} u_{S_i(j)},
	\label{eq:operatorApprox}
\end{equation}
based on the function values $u$ in stencil nodes and weights $w$ that are pre-computed by demanding the exactness of Eq. \eqref{eq:operatorApprox} for a set of basis functions. Note that the same general framework used for differential operators can also be used with $\mathcal{L} = 1$ for approximation of function values. We use radial basis functions 
\begin{equation}
	\label{eq:rbfScaling}
	\Phi(i, j) = \Phi \left( \frac{\lVert \vec{x}_j - \vec{x}_i \rVert_2}{\delta_i} \right),
\end{equation}
with a local scaling factor $\delta_i$ that decouples the approximation from the choice of the coordinate system (especially important for RBF with a shape parameter\cite{lehto2017rbf}). It can be set to an arbitrary local measure of distance, e.g., the distance between the central node and its closest neighbor.
Replacing $u$ with the RBF in the exact form of Eq. \eqref{eq:operatorApprox} leads to a local linear system $\vec{A} \vec{w}_i = \vec{b}$
\begin{gather}
	\label{eq:nonAugmentedSystem}
	\begin{split}
		\begin{bmatrix}
			\Phi(S_i(1), S_i(1)) & \cdots  & \Phi(S_i(1), S_i(N_L)) \\
			\vdots & \ddots & \vdots \\
			\Phi(S_i(N_L), S_i(1)) & \cdots & \Phi(S_i(N_L), S_i(N_L)) \\
		\end{bmatrix}
		\begin{bmatrix}
			w_{i, 1} \\ \vdots \\ w_{i, N_L}
		\end{bmatrix}
		=
		\begin{bmatrix}
			(\mathcal{L}\Phi)(i, S_i(1)) \\
			\vdots \\
			(\mathcal{L}\Phi)(i, S_i(N_L))
		\end{bmatrix},
	\end{split}
\end{gather}
for each stencil with the solution providing approximation weights $\vec{w}_i$. The right 
hand side vector $\vec{b}$ is formed by applying the linear operator $\mathcal{L}$ to 
the basis function and evaluating the result with an argument analogous to Eq.~\eqref{eq:rbfScaling}.

We use an RBF with no shape parameter, the polyharmonic spline (PHS)
\begin{equation}
	\Phi(r) = r^{k},
\end{equation}
with odd order $k$, to avoid additional problems with parameter tuning. Local matrices constructed with polyharmonic RBF are only conditionally positive definite and need to be augmented with monomials \cite{flyer2016polynomials1, slak2020thesis}. The system is expanded with $N_p = \binom{m + D}{m}$ monomials\footnote{The 10 monomials in $D=3$ case with $m=2$ would be $p = \{1, x, y, z, x^2, xy, xz , y^2, yz, z^2\}$.} $p_l$, where $m$ denotes the monomial order. The monomials are scaled similarly to RBFs in Eq.~\eqref{eq:rbfScaling}
\begin{equation}
	p_l(i, j) = p_l \left( \frac{\vec{x}_j - \vec{x}_i}{\delta_i} 
	\right).
\end{equation}
The linear system from Eq.~\eqref{eq:nonAugmentedSystem} is augmented with monomials
\begin{equation}\label{eq:augmentedSystem}
	\begin{gathered}
		\begin{bmatrix}
			\vec{A} & \vec{P} \\
			\vec{P}^T & 0 \\
		\end{bmatrix}
		\begin{bmatrix}
			\vec{w}_i \\ \vec{\lambda}
		\end{bmatrix}
		=
		\begin{bmatrix}
			\vec{b} \\
			\vec{c} \\
		\end{bmatrix},
		\\
		\vec{P} = \begin{bmatrix}
			p_1(S_i(1), S_i(1)) & \cdots  & p_{N_p}(S_i(1), S_i(1))\\
			\vdots & \ddots & \vdots \\
			p_1(S_i(N_L), S_i(1)) & \cdots & p_{N_p}(S_i(N_L), S_i(1))\\
		\end{bmatrix}
		, \qquad
		\vec{c} = \begin{bmatrix}
			(\mathcal{L} p_1)(S_i(1), S_i(1)) \\
			\vdots \\
			(\mathcal{L} p_{N_p})(S_i(1), S_i(1))
		\end{bmatrix},
	\end{gathered}
\end{equation}
with the additional weights $\vec{\lambda}$ treated as Lagrange multipliers 
and discarded after computation. 

Augmentation with an order of at least $m = \frac{k - 1}{2}$ is required to guarantee the positive definiteness for a PHS with order $k$. Higher orders provide better convergence characteristics \cite{jancic2021monomial} at the cost of increased computational complexity, since the required 
stencil size is $N_L \ge N_p$, with $N_L > 2N_p$ as the often recommended value \cite{Bayona2017}. We use a slightly larger stencil size $N_L = 25$ than the recommended minimum 20 based on observations in \ref{ssec:refinement}. In the remainder of the paper, we use approximations with $k=3$ and $m=2$ unless otherwise specified.

\subsection{Meshless Lattice Boltzmann Method}\label{ssec:MLBM_method}

In the present study we use the meshless Lattice Boltzmann Method the principles of which are outlined in \cite{Lin2019,Strzelczyk2024}. It solves the discrete velocity Boltzmann equation\cite{Succi2018}:
\begin{equation}\label{eq:LBM_streaming}
    f_k\left(t+1,\bsym{x}\right) = f^\text{post}_k\left(t,\bsym{x}+\boldsymbol{e}_{k'}\right), \quad k=0,1,\dots,q-1
\end{equation}
in a semi-Lagrangian way. In the above equation, $f_k$ is the $k$-th distribution function, $\bsym{e}_k=-\bsym{e}_{k'}$ is the $k$-th streaming direction and its opposite, respectively, and superscript "post" denotes post-collision distribution. We implement compressible D3Q15 model ($q\!=\!15$) with the following set of discrete streaming directions:
\begin{equation}
    \begin{array}{cc}
         k&\bsym{e}_k\\
         \hline
         0 & (0,0,0)\\
         1 & (1,0,0)\\
         3 & (0,1,0)\\
         5 & (0,0,1)\\      
    \end{array}\quad
    \begin{array}{cc}
         k&\bsym{e}_k\\
         \hline
         7 & (1,1,1)\\
         9 & (1,1,-1)\\
         11 & (1,-1,1)\\
         13 & (1,-1,-1)\\      
    \end{array},\quad
    \bsym{e}_k = - \bsym{e}_{k-1} \text{ for } k=2,4,\dots,14
\end{equation}
In the present work we use the two relaxation time kernel (TRT)\cite{Ginzburg2008,Krueger2016}. As we are concerned with the Stokes flow, the stability provided by TRT collision is satisfactory. We note that should higher velocities occur in the system, one can choose from more stable kernels, such as regularized two-relaxation time model~\cite{Yu2024,Yu2025}.
The post-collision distributions $f^\text{post}_k$ in TRT-LBM have the form of
\begin{equation}\label{eq:LBM_collision}
    f^\text{post}_k(t,\bsym{x})=f_k(t,\bsym{x}) - \frac{1}{\tau^+}\left(f_k^+(t,\bsym{x})-f_k^{\text{eq},+}(t,\bsym{x})\right) - \frac{1}{\tau^-}\left(f_k^-(t,\bsym{x})-f_k^{\text{eq},-}(t,\bsym{x})\right) + F_k
\end{equation}
where $\tau^+$ and $\tau^-$ are symmetric and anti-symmetric relaxation parameters related by
\begin{equation}
    \Lambda = (\tau^+ - 0.5)(\tau^- - 0.5),
\end{equation}
with the values of $\Lambda=1$ and $\tau^+=1$ used in the present study. $f_k^+$ and $f_k^-$ are the symmetric and anti-symmetric components of the $k$-the distribution function and $f_k^{\text{eq},+}$, $f_k^{\text{eq},-}$ are symmetric and anti-symmetric components of the $k$-th equilibrium distribution function
\begin{equation}
    \begin{aligned}
        f_k^+ = \frac{1}{2}(f_k+f_{k'}),& \quad f_k^{\text{eq},+} = \frac{1}{2}(f_k^\text{eq}+f_{k'}^\text{eq}),\\[.5ex]
        f_k^- = \frac{1}{2}(f_k-f_{k'}),& \quad f_k^{\text{eq},-} = \frac{1}{2}(f_k^\text{eq}-f_{k'}^\text{eq}),\\[.5ex]
    \end{aligned}
\end{equation}
The relaxation time $\tau^+$ define the kinematic viscosity of the considered fluid in each LBM model
\begin{equation}\label{eq:lbm_viscosity}
    \nu_{lb} = c_s^2\left(\tau^+ - \frac{1}{2}\right),
\end{equation}
where the lattice speed of sound is $c_s=1/\sqrt{3}$ and the subscript $lb$ denotes a macroscopic quantity in LBM model units (non-dimensional), to distinguish it from the same quantity in physical units which will be denoted without any subscript. We use the second-order discretization of the equilibrium distribution function
\begin{equation}\label{eq:feq}
	f^\text{eq}_k = \rho_{lb}\omega_k\left[1+\frac{\boldsymbol{e}_k\cdot \boldsymbol{v}_{lb}}{c_s^2} + \frac{(\boldsymbol{e}_k\cdot \boldsymbol{v}_{lb})^2}{2c_s^4} - \frac{\boldsymbol{v}_{lb}^2}{2c_s^2}\right],
\end{equation}
where lattice weights $\omega_k$ have the values of:
\begin{equation}
    \omega_k=
    \begin{cases}
        2/9, \> k=0 \\
        1/9, \> k=1,\dots,6 \\
        1/72, \> k=7,\dots,14
    \end{cases},
\end{equation}
and $\rho_{lb}$ and $\bsym{v}_{lb}$ are local fluid macroscopic density and velocity vector, respectively:
\begin{equation}\label{eq:macro_var}
	\renewcommand{\arraystretch}{2.5}
	\begin{aligned}
		\rho_{lb} = &\>\sum\limits_{k=0}^{q-1} f_k \\
		\bsym{v}_{lb} = &\>\frac{1}{\rho_{lb}}\sum\limits_{k=0}^{q-1} f_k \bsym{e}_k
	\end{aligned}
\end{equation}

To obtain the relative pressure we use the ideal gas equation of state with the deviation of the density from its mean:
\begin{equation}
    p_{lb}=\left(\rho_{lb}-\langle \rho_{lb} \rangle\right)c_s^2.
\end{equation}

We implement acceleration using first-order discretization in velocity space:
\begin{equation}\label{eq:force_lbm}
    F_k = \omega_k \frac{\bsym{e}_k \cdot \bsym{F}_{lb}}{c_s^2}
\end{equation}
where $\bsym{F}_{lb}=\rho_{lb} \bsym{g_{lb}}$ is the body force defined in terms of acceleration $\bsym{g_{lb}}$.

The conversion from LBM units to physical units is achieved by the multiplication of a quantity by a suitable conversion factor:
\begin{equation}\label{eq:conversion_factors}
    \begin{array}{lrlr}
        L = \delta x L_{lb},&\quad\text{length},&\hspace{2cm}\nu = \frac{\delta x^2}{\delta t} \nu_{lb},& \quad\text{viscosity},\\[.5ex]
        T = \delta t T_{lb},& \quad\text{time},&\hspace{2cm} g = \frac{\delta x}{\delta t^2} g_{lb},& \quad\text{acceleration},\\[.5ex]
        v = \frac{\delta x}{\delta t} v_{lb},& \quad\text{velocity},& \hspace{2cm}\rho = \rho_{ref} \rho_{lb},& \quad\text{density},\\[.5ex]
        &&\hspace{2cm}p = \rho_{ref}\frac{\delta x^2}{\delta t^2}p_{lb},& \quad \text{pressure}\\[.5ex]
    \end{array}
\end{equation}
where $\delta x$, $\delta t$, $\rho_{ref}$ are the streaming distance length, the timestep length and the reference density, respectively, all in physical units. By virtue of Chapman-Enskog expansion \cite{Krueger2016}, the macroscopic velocities and pressure obtained from the solution of the discrete velocity Boltzmann equation are the solution to the weakly compressible Navier-Stokes equation.

\begin{figure}[!ht]
\centering
\includegraphics[width=.75\linewidth]{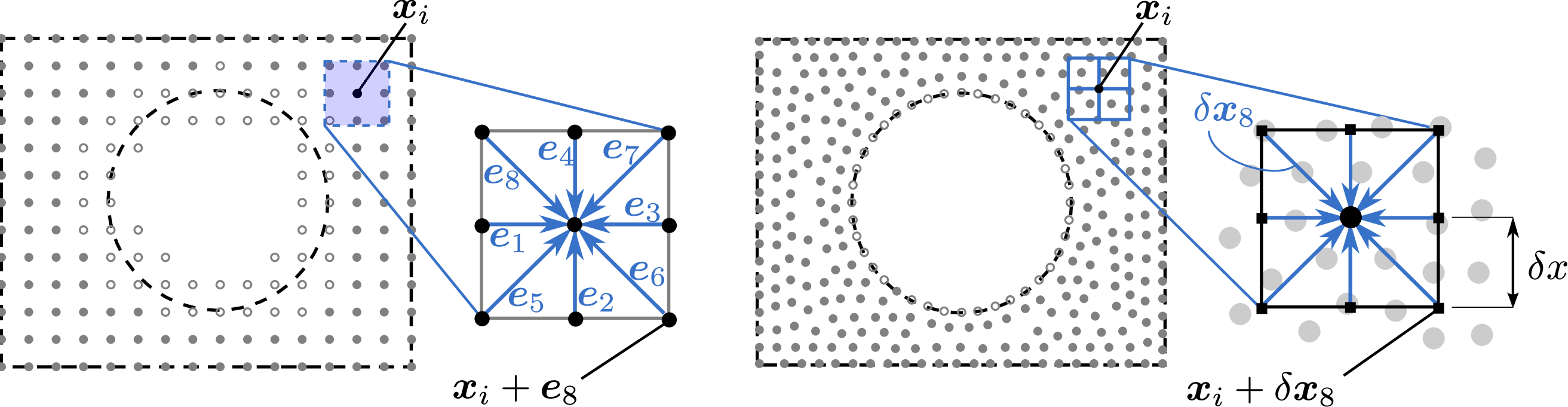}
\caption{Graphical interpretation of differences between the space discretization of a circle in standard (lattice-based) LBM (\textit{left}) and MLBM (\textit{right}). The open symbols in LBM case represent the nodes whose links are cut by the circle's boundary. In MLBM the open symbols represent the boundary nodes lying exactly on the circle's surface.}
\label{fig:mlbm_discretization}
\end{figure}

Since the flow domain is discretized with a set of scattered nodes, it is convenient to reformulate the streaming step (Eq.~\eqref{eq:LBM_streaming}) into physical units and speak of the streaming distances $\delta \bsym{x}_k=\bsym{e}_k \delta x$ rather than streaming directions (see Fig.~\ref{fig:mlbm_discretization}). At the same time, the collision step (Eq.~\eqref{eq:LBM_collision}) is still solved non-dimensionalized. In this way, the positions of the departure (Lagrangian) nodes for $f_k^\text{post}$ can be related to the meshless space discretization (Eulerian) nodes as $\bsym{x}+\delta\bsym{x}_{k'}$. In contrast to the lattice-based LBM, they do not need to coincide with the Eulerian nodes. Unless otherwise stated, in the presented MLBM setups we used the streaming distance equal to half of the minimal internodal distance, i.e. $\delta x = h_\text{min}/2$ (compare with Eq.~\eqref{eq:refinement_function}). Along with the relaxation time $\tau$ and the physical kinematic viscosity $\nu$ it determines the timestep length $\delta t$ according to the viscosity conversion from Eq.~\eqref{eq:conversion_factors}\footnote{The actual values of $\delta t$ that we use range from $1.04\cdot 10^{-6}$ to $1.67\cdot 10^{-5}$.}.

To obtain the value of the post-collision distribution function in such a case, we use a method of meshless interpolation described in Sec.~\ref{ssec:meshless} ($\mathcal{L}=1$ in Eq.~\eqref{eq:operatorApprox}).

In MLBM the no-slip condition is achieved through the interpolated bounce-back~\cite{Pan2006} applied to the boundary nodes. As the discretization nodes lie exactly on the boundary, the formula for the bounce-back simplifies to:
\begin{equation}\label{eq:interpolated_bounceback_mlbm}
    f_k(t+\delta t,\bsym{x})=f_{k'}(t+\delta t,\bsym{x}).
\end{equation}

The initial conditions for all simulations presented in this work were equilibrium populations (Eq.~\eqref{eq:feq}) parametrized with zero macroscopic velocities $\bsym{v}_{lb}$ and unit density $\rho_{lb}$. The Mach number did not exceed $2\cdot10^{-4}$ in all simulations in the steady state.

\subsection{Meshless Navier-Stokes solver}\label{ssec:MNS_method}

One of the main problems in directly solving the Eq.~\eqref{eq:NSeq} lies in implementing the pressure-velocity coupling procedure that ensure continuity. We use one of the fundamental approaches, the artificial compressibility method (ACM)\cite{trojak2022acm, kosec2018localIrregularFlow}, first introduced by Chorin~\cite{chorin1967acm} in 1967, that relies on transiently introducing a slight compressibility into the system. The main benefit of this method from the computational standpoint is that it avoids solving the global pressure Poisson equation required by pressure projection\cite{ma2016projection}, the other main class of pressure-velocity coupling methods, thus allowing for perfect parallelization within a single time step. Additionally, the selection of ACM for pressure-velocity coupling is motivated by its innate similarities to LBM~\cite{Ohwada2011ACMvsLBM}. In this paper we resorted to a relatively simple implementation of the ACM, but advanced versions, like the entropically damped artificial compressibility~\cite{clausen2013EDAC}, could resolve some of the issues with oscillatory pressure in the stagnant regions that we observed in the results.

We use the explicit Euler method for the temporal discretization of the Navier-Stokes equation
\begin{equation}
    \vec{v'} = \vec{v} + \delta t \left( \nu \nabla^2 \vec{v} - \vec{v} \cdot \nabla\vec{v} + \vec{g} \right), \label{eq:MNS_intermediate}
\end{equation}
where $\delta t$ is the time step, $\mu$ the the viscosity and $\rho$ the density. The time step\footnote{The actual values of $\delta t$ that we range from $1.25 \cdot 10^{-6}$ to $2 \cdot 10^{-5}$.} $\delta t = 0.1\frac{h_\text{min}^2}{2 \nu}$ is set as a function of the smallest inter-nodal distance $h_\text{min}$. Note that the pressure term is omitted while calculating the intermediate velocity $\vec{v}'$. This is because the subsequent pressure-velocity coupling
\begin{align}
    p & \leftarrow p - \delta t C^2  \rho  (\nabla \cdot \vec{v}), \label{eq:MNS_pressure} \\
    \vec{v} &\leftarrow \vec{v'} - \frac{\delta t}{\rho} \nabla p, \label{eq:MNS_velocity}
\end{align}
is done iteratively. The magnitude of the AC is determined with the artificial speed of sound $C$~\cite{rahman2008ACM}
\begin{equation}
    C = \beta \max(\max_i(\lVert\vec{v_i}\rVert_2), \lVert\vec{v}_{ref}\rVert_2), \label{eq:MNS_C}
\end{equation}

where $\beta$ is the compressibility parameter, and $\vec{v}_{ref}$ a reference velocity that prevents instabilities due to $C$ reaching zero in stagnation points. We use $\beta = 10$ for all computations presented in this paper.
If we sought a time-accurate solution the pressure-velocity coupling iteration would have to be iterated until the maximum divergence of the velocity field dropped below a pre-determined threshold, but since we are dealing with a steady state problem, the number of iterations can be limited. We chose to perform $n_p=3$ pressure correction iterations for each velocity iteration as a reasonable compromise that empirically provided the fastest convergence to a steady state.

The non-slip boundary condition (BC) is enforced by setting boundary node values according to the Dirichlet BC $\vec{v} = 0$ for velocity and the Neumann BC $\frac{\partial p}{\partial \vec{n}} = 0$ for pressure.

\subsection{Case definition}

We consider a flow through a three-dimensional, infinite, periodic, simple cubic (SC) array of spheres. All the spheres in the domain have the same radius and we consider cases with radii ranging from $r=0.04$ to $r=0.69$. This gives the porosity range $\varphi \in [0.05,0.9997]$ and a natural transition from the overlapping to the diluted regime of the system. The flow is forced with a constant acceleration $\bsym{g}=[0.1,0,0]$. In actual computations, it is thus sufficient to reduce the computational domain to the fluid-occupied part of a single periodic cell with the side length $d=1$. The no-slip boundary condition is imposed on the spheres' walls and periodic boundary conditions are imposed in all three directions via the periodic search of stencils' members.

\begin{figure}[!ht]
\centering
\includegraphics[width=.5\linewidth]{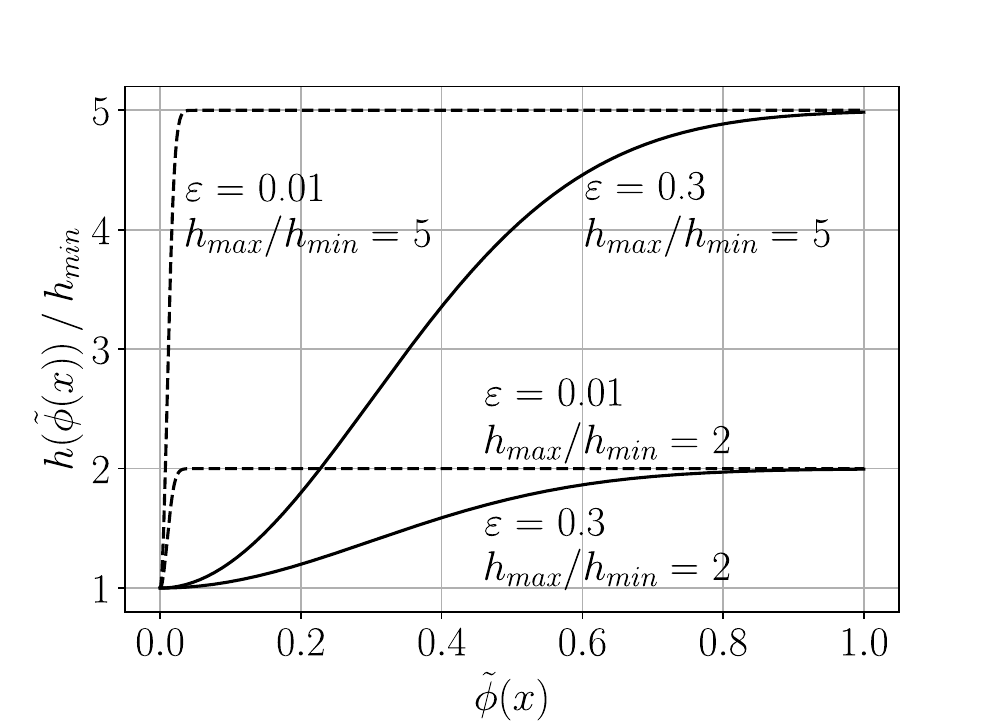}
\caption{The average distance between points as a function of the normalized distance function value for various shape parameters and $h_\text{max} / h_\text{min}$ ratios.}
\label{fig:hFunctions}
\end{figure}

\begin{figure}[!ht]
\centering
\includegraphics[width=.24\linewidth]{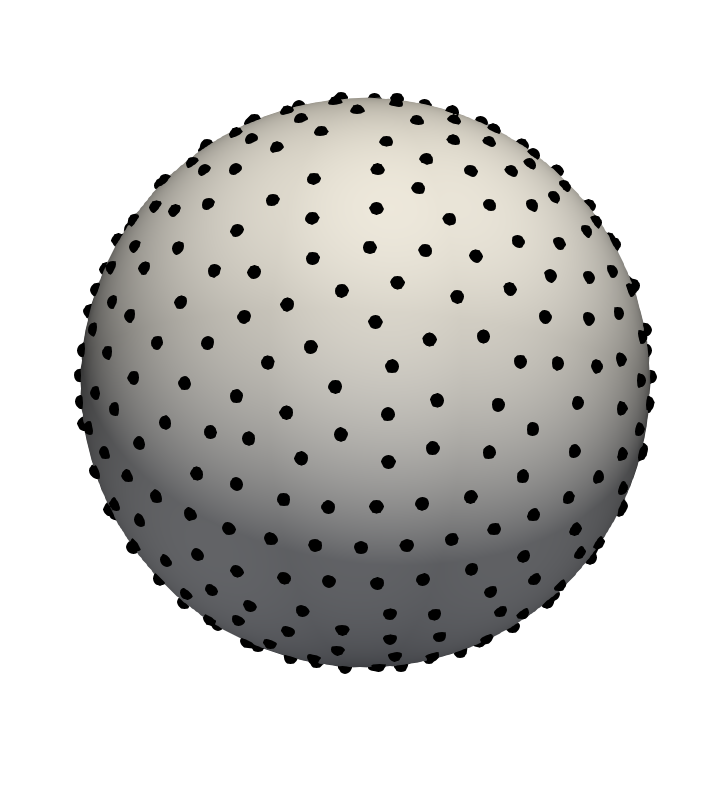}
\includegraphics[width=.24\linewidth]{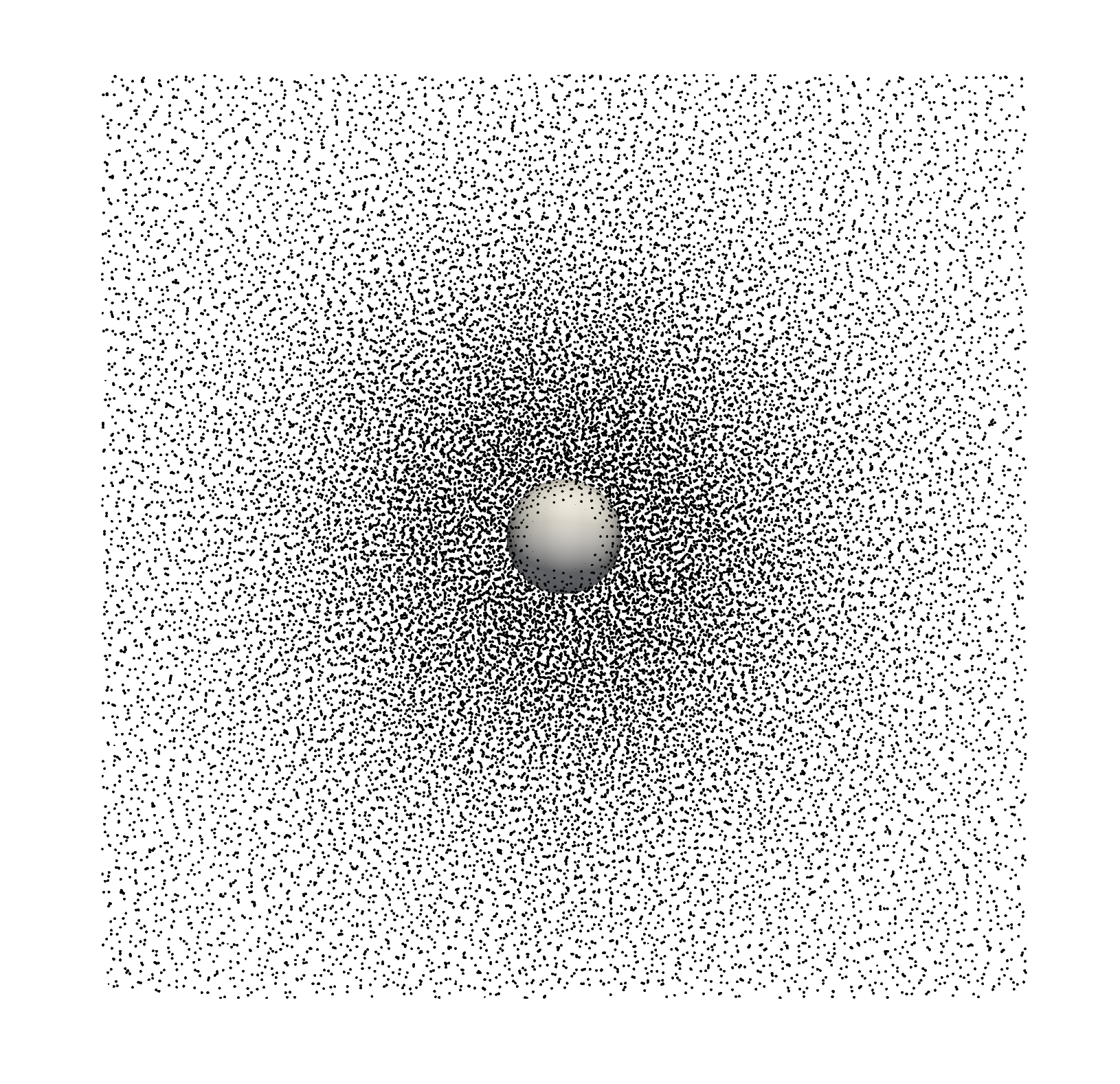}
\includegraphics[width=.24\linewidth]{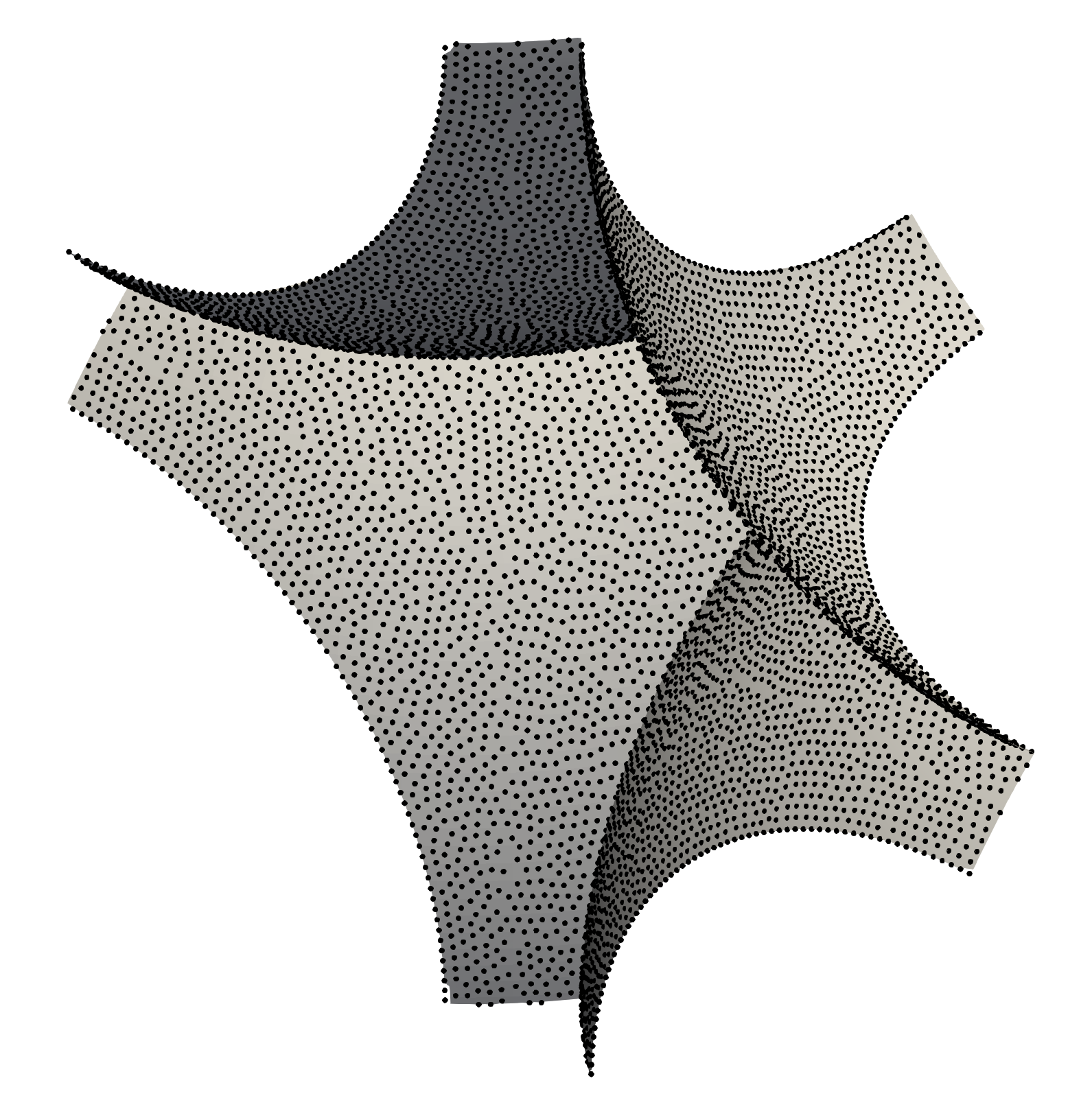}
\includegraphics[width=.24\linewidth]{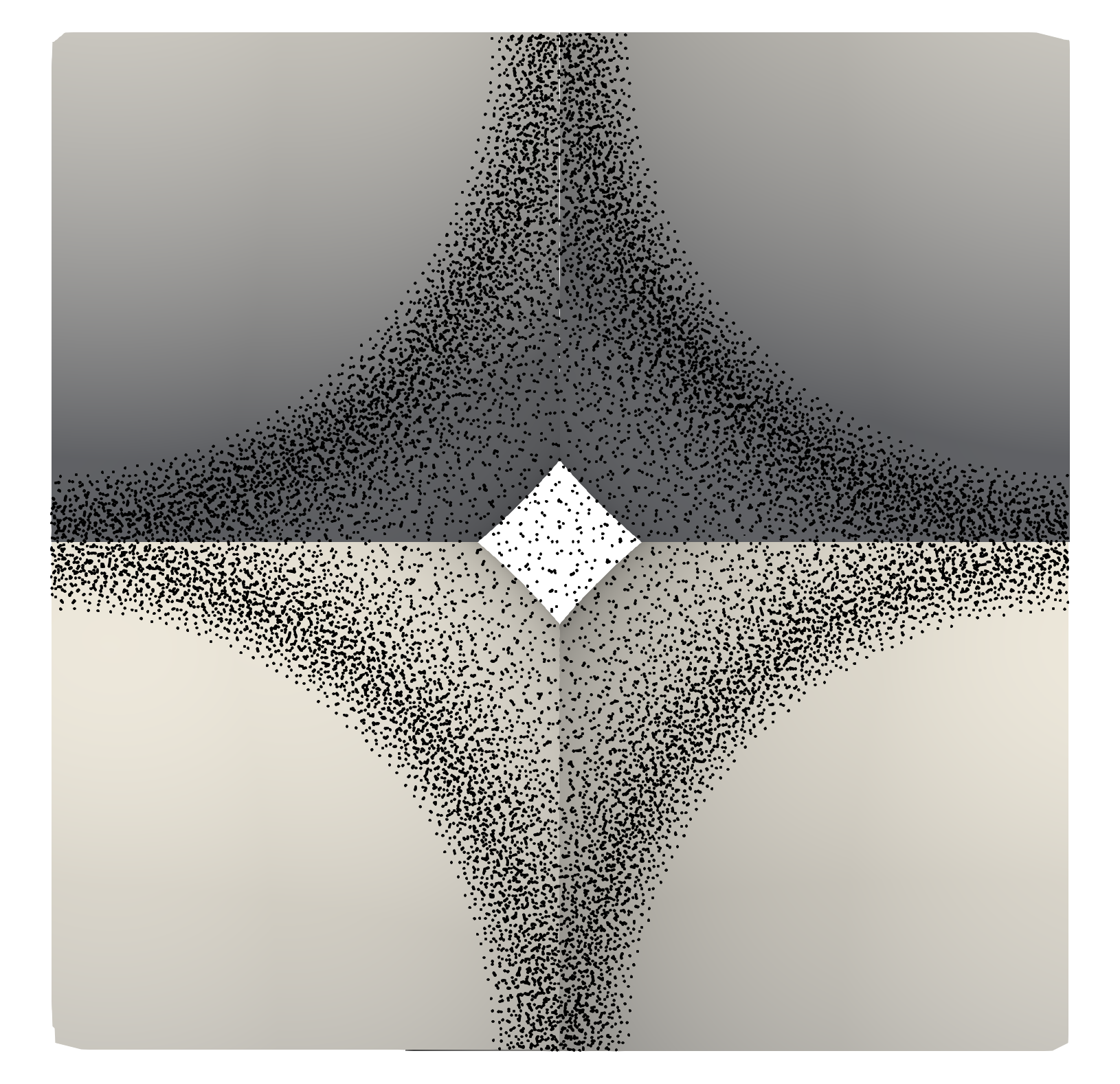}
\caption{Meshless space discretizations used in the calculations. \textit{Two leftmost pictures:} $r=0.062$, \textit{two rightmost pictures:} $r=0.6526$. Each pair consists of the visualization of the boundary discretization and the subset of nodes with $y \in [0.45;0.55]$ viewed along +$y$ direction. The spheres' boundaries are rendered as grey surfaces.}
\label{fig:discretizations}
\end{figure}

The discretization is refined locally by decreasing the internodal distance towards the spheres' surface. We implement the refinement by specifying the function of the target distance between the nodes:
\begin{equation}
\label{eq:refinement_function}
    h(\bsym{x})=h_\text{max} - \exp\left(-\frac{\tilde\phi^2(\bsym{x})}{2\varepsilon^2}\>\right)(h_\text{max}-h_\text{min}), \quad \tilde\phi(\bsym{x}) = \phi_\text{sdf}(\bsym{x})\left(\sqrt{3}/2-r\right)
\end{equation}
where $\phi_\text{sdf}(\bsym{x})$ is the distance of the point $\bsym{x}$ to the closest spherical obstacle, $h_\text{min}$ and $h_\text{max}$ are minimum and maximum values of the target distance, respectively and $\varepsilon$ is the shape parameter. Note that $h(\bsym{x})$ reaches the minimum on the spheres' surface ($\phi_\text{sdf}(\bsym{x})=0$) and maximum away from it. The shape of the refinement function for a selection of parameters is shown in (Fig.~\ref{fig:hFunctions}). Unless otherwise stated we use the values of $h_\text{min}=0.5h_\text{max}$ and $\varepsilon=0.295$. Example visualizations of the discretizations are shown in Fig.~\ref{fig:discretizations}.

To allow for a reliable comparison between MLBM and meshless Navier-Stokes results we use the stopping criterion based on the relative change of permeability $k/d^2$
\begin{equation}\label{eq:stopCrit}
    |\Delta|_{k/d^2} = \frac{1}{\left(k/d^2\right)(t)}\>\frac{\left|\left(k/d^2\right)(t) -\left(k/d^2\right)(t - \Delta t)\right|}{\Delta t}
\end{equation}
calculated every $10^4$ timesteps with $\Delta t = 10^4 \delta t$ and we stop the simulations when $|\Delta|_{k/d^2}<10^{-2}$. The permeability is defined as:
\begin{equation}\label{eq:k_definition}
    k/d^2 = \frac{q \nu}{|\bsym{g}|d^2}, \quad
    q = \frac{1}{d^2} \cdot \frac{1}{5} \sum\limits_{i=1}^5 \int\limits_{P_i} d\!A \> v_0(\bsym{x})
\end{equation}
where $q$ is the mean $x$-component of the velocity averaged over five cross-sections $P_i = \left\{ (x, y, z) \in \Omega : x \in \{0,0.2,0.4,0.6,0.8\}\right\}$.

In both methods, for the calculation of surface forces acting on the spheres we use the stress tensor $\sigma_{ij}$ at boundary nodes:
\begin{equation}\label{eq:stress_tensor_force}
    \sigma_{ij} = -p\delta_{ij} + \rho\nu\left(
    \frac{\partial v_i}{\partial x_j}+\frac{\partial v_j}{\partial x_i}
    \right)
\end{equation}
The integration of the stresses over the spheres' surface assumes equal area $\Delta A$ assigned to each boundary node:
\begin{equation}
    \Delta A = A / N_b
\end{equation}
where $N_b$ is the number of boundary nodes and $A$ is the area of the obstacles in a unit cell. To obtain the hydrodynamic force on the obstacle we sum:
\begin{equation}\label{eq:integration_force}
    F_{H,i}=\Delta A\sum\limits_{n\in I_b} \left(\sigma_{ij} \hat{n}_j\right)_{\bsym{x}_n}
\end{equation}
where $\hat{n}_i$ is a local normal vector and $I_b$ is the set of boundary nodes indices.

\section{Results and Discussion}
\subsection{Analysis of the velocity and stresses fields}\label{ssec:flow_fields}

\begin{figure}[!ht]
\centering
\includegraphics[height=.42\linewidth]{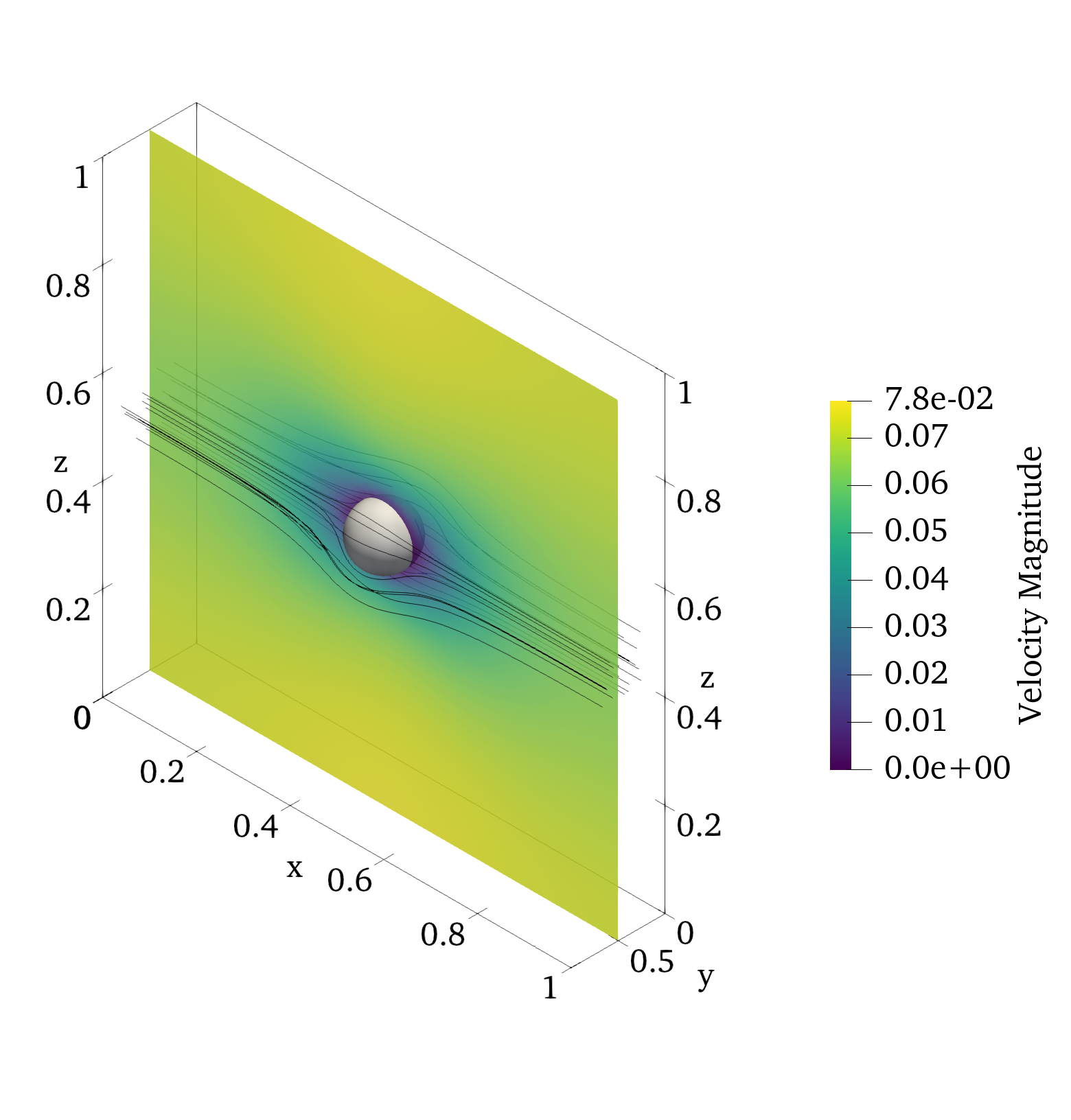}
\hspace{0.5cm}
\includegraphics[height=.42\linewidth]{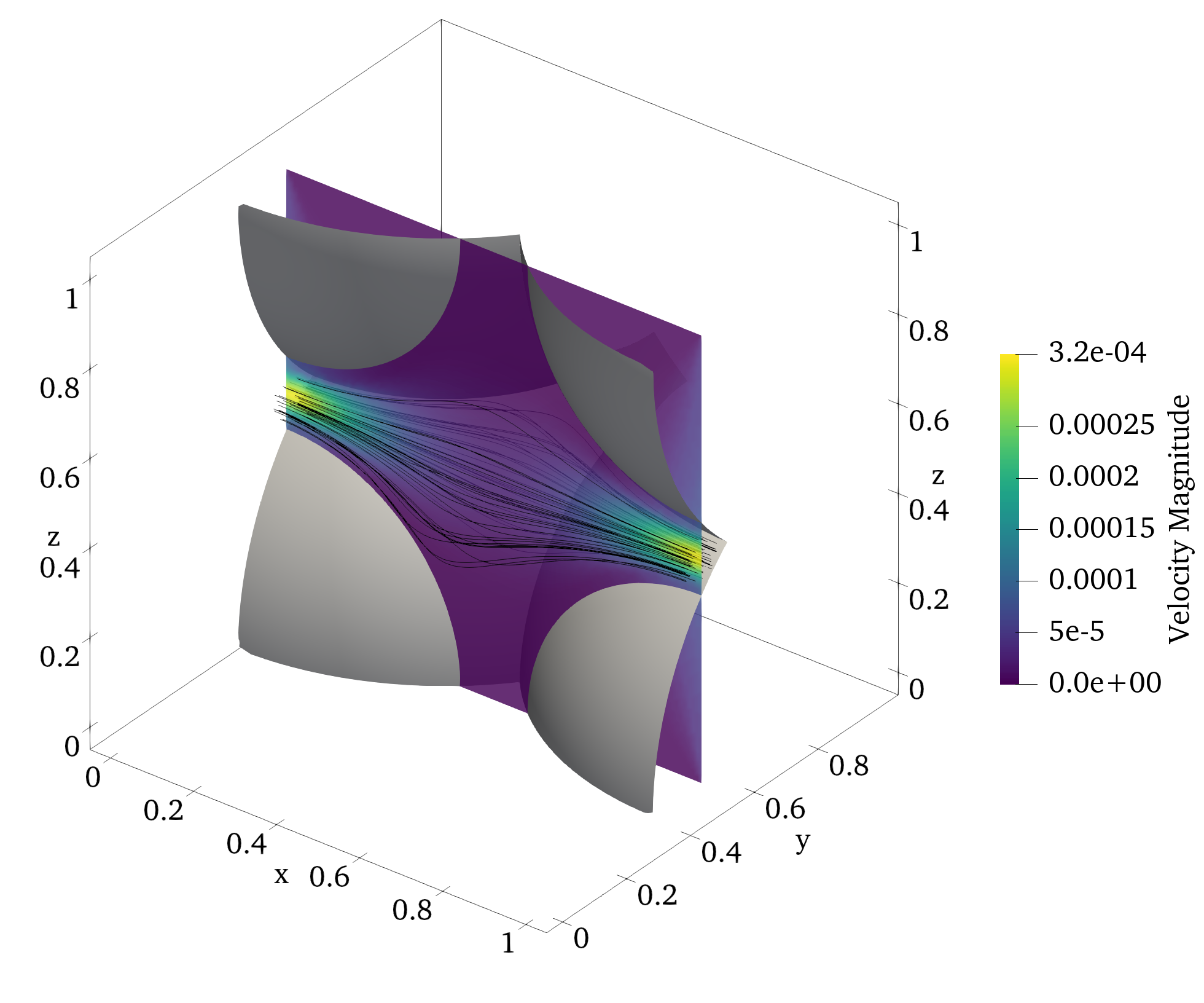}
\caption{Streamlines of the velocity field and a cross section of the velocity magnitude field for $r=0.062$ (\textit{left}) and $r=0.6526$ (\textit{right}) case.}
\label{fig:flow_viz}
\end{figure}

We first present the visualizations of the velocity fields obtained for diluted and overlapping systems in Fig.~\ref{fig:flow_viz}. For the small sphere case, we observe an undisturbed flow away from the obstacle and a boundary layer near the sphere's surface. For the large sphere case, a strong channelization is visible with a considerable portion of the fluid volume excluded from the percolation where recirculation occurs. Also, the value of the maximum velocity magnitude differs by two orders of magnitude between the two cases.

\begin{figure}[!ht]
\centering
\includegraphics[height=.3\linewidth]{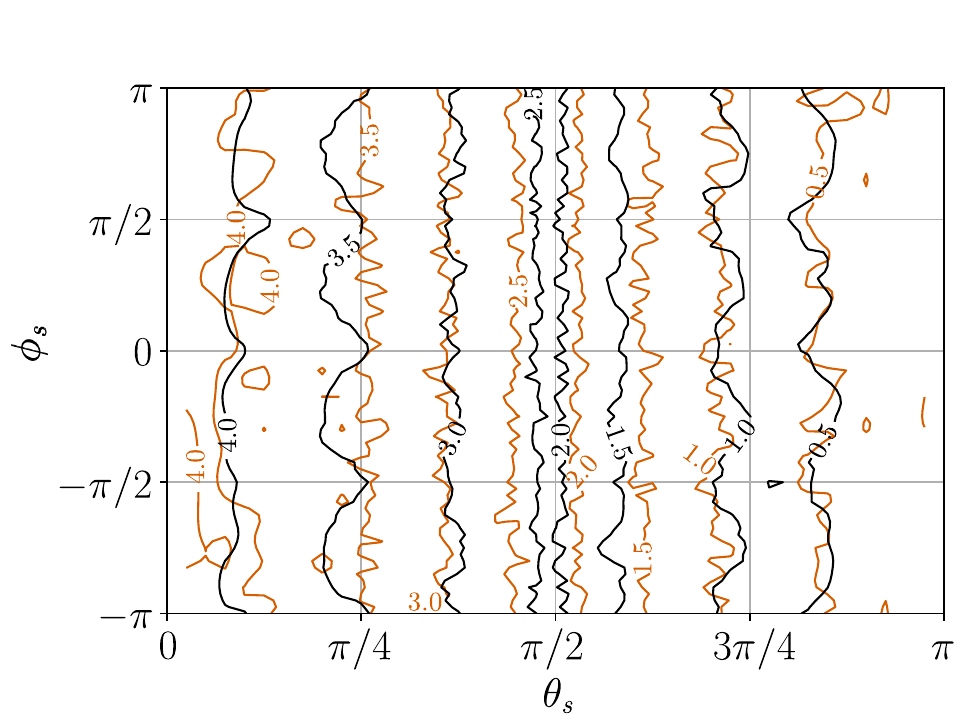}
\includegraphics[height=.3\linewidth]{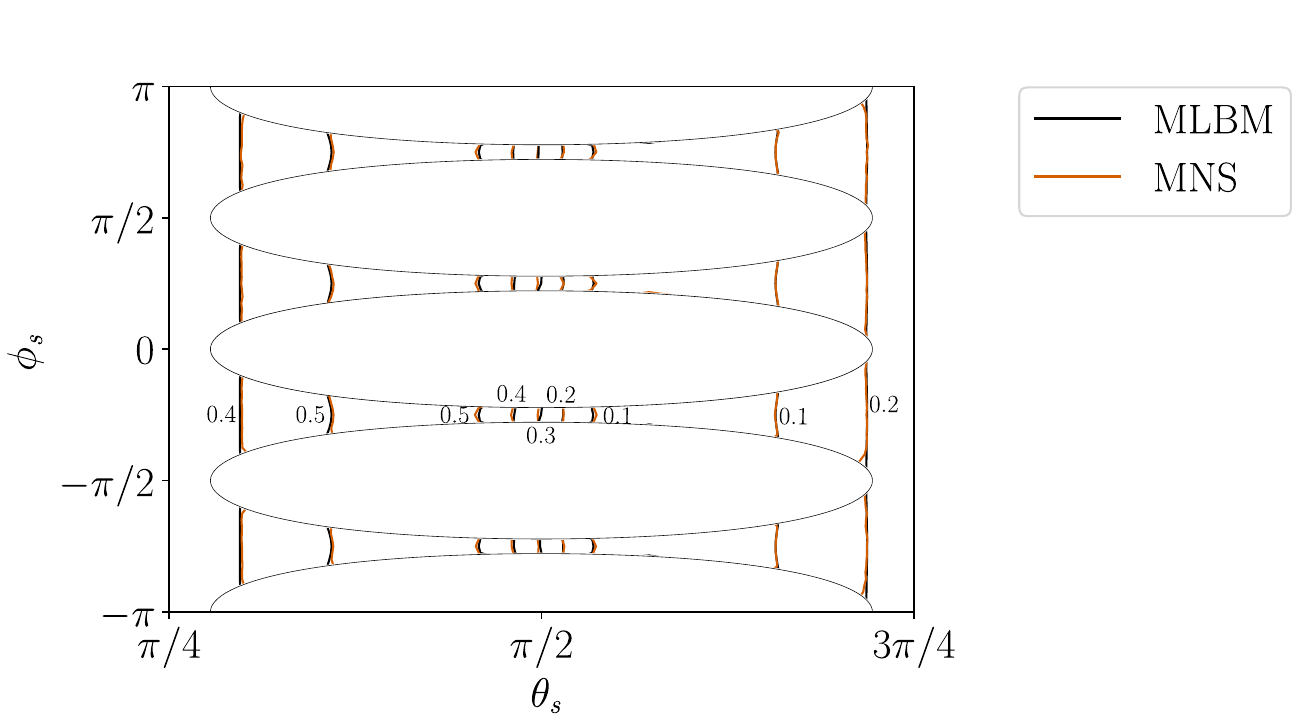}
\includegraphics[height=.3\linewidth]{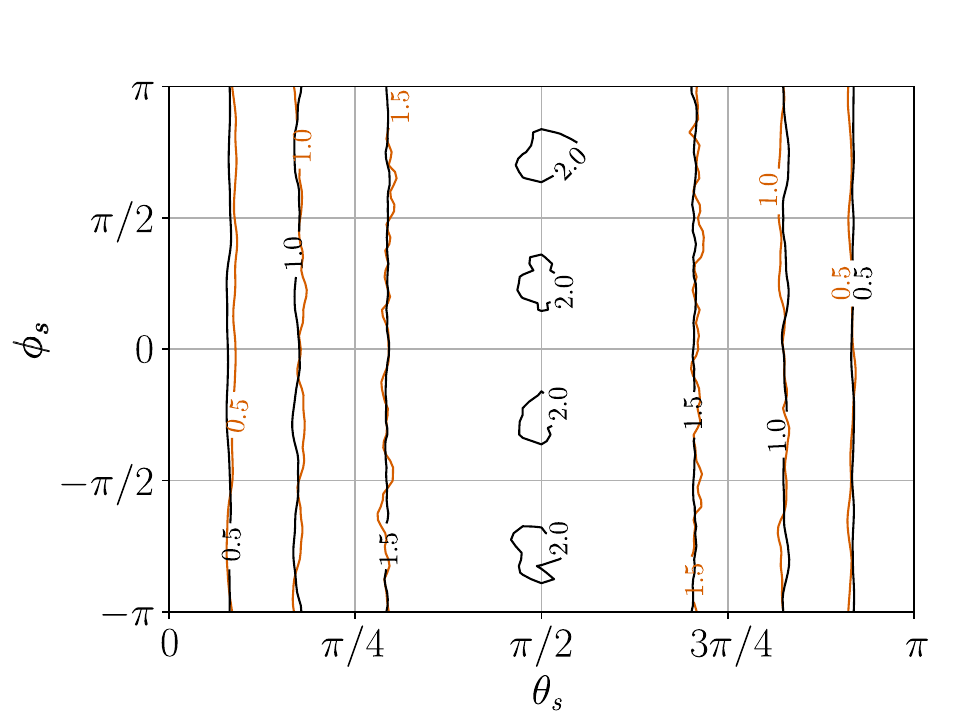}
\includegraphics[height=.3\linewidth]{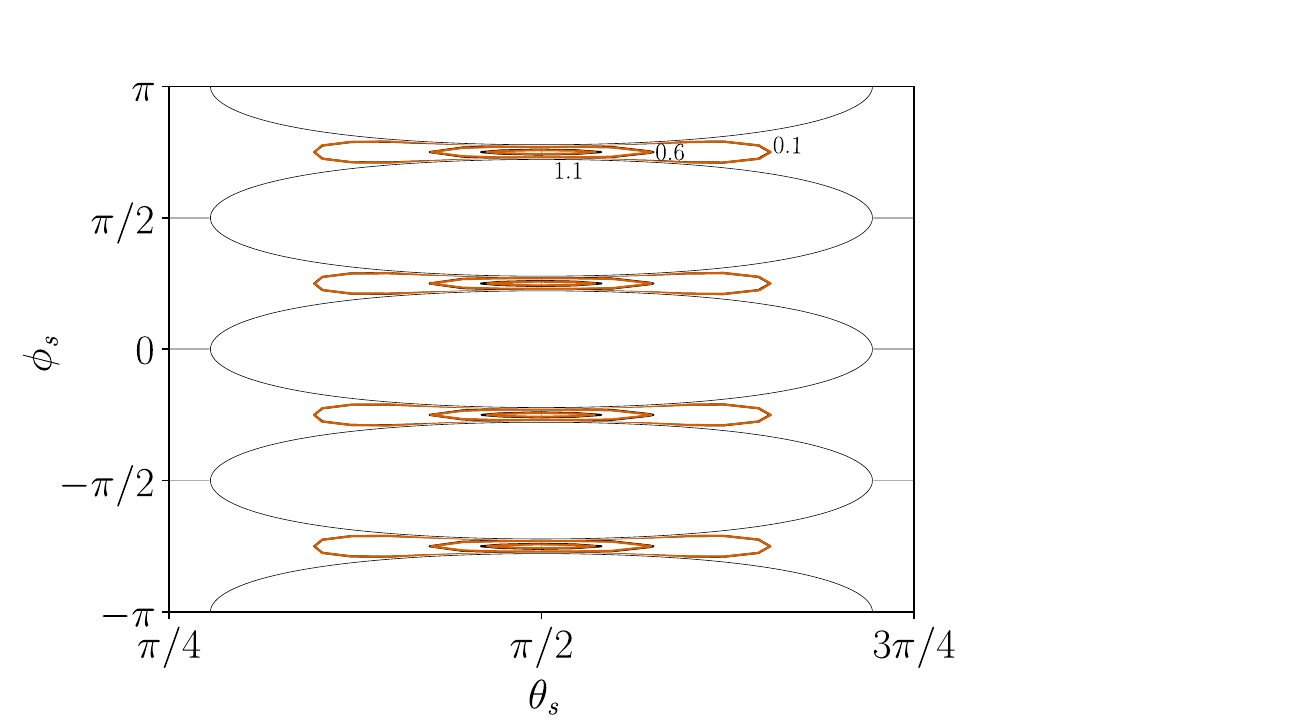}
\caption{Pressure (\textit{top}) and tangent force (Eq.~\eqref{eq:integration_force}) magnitude (\textit{bottom}) contours on the sphere's surface for $r=0.062$ (\textit{left}) and $r=0.6526$ (\textit{right}) obtained with MLBM and MNS. The contours are presented in a spherical coordinate system with the origin at the sphere's center. $\theta_s$ is the polar angle measured from the negative direction of the $x$-axis and $\phi_s$ is the azimuthal angle measured from the positive direction of the $z$-axis. The blank ellipses in the right column result from the intersection between the obstacle and the periodic cubic cell.}
\label{fig:pressure_cp}
\end{figure}

Initially, we performed a set of simulations for $|\bsym{g}|=49$ as indicated in \cite{Holmes2011}.
We noticed, however, that both results obtained with MLBM and MNS for $|\bsym{g}|=49$ exhibit an asymmetry about the $x=0.5$ plane in the $x$-component of the velocity field at high porosities (data not shown). It can be associated with inertial effects in the flow. If we define the Reynolds number as:
\begin{equation}
    Re = \frac{2rq}{\nu},
\end{equation}
we find that for $g=49$ and $r=0.062$ the Reynolds number is approximately $4$. This is already when inertial effects appear in the flow past a single sphere\cite{Almedeij2008}. This motivated us to decrease the acceleration to $|\bsym{g}|=0.1$ to ensure Darcy regime in the whole range of porosities. Now, for the lower acceleration, the Reynolds number for the same system is on the order of $10^{-2}$.

Fig.~\ref{fig:pressure_cp} shows the isocontours of pressure and tangent force magnitude on the sphere's surface for $r=0.062$ and $r=0.6526$. Results obtained with both methods are compliant with each other in terms of the values of the quantities.  The most pronounced difference between the two methods is visible in the tangent force isocontours for $r=0.062$ case, where MNS records values $\ge 2$, not present in MLBM solution. Note that the MNS solution exhibits more fluctuations in pressure isocontours, which is consistent with the known ACM problems where pressure oscillations appear in the low Reynolds regime\cite{clausen2013EDAC}. Nevertheless, this level of compliance between the methods is satisfactory, especially as MLBM and MNS use vastly different strategies to calculate the pressure (see Sec.~\ref{sec:methods}). Moreover, the use of boundary-compliant discretization in MLBM allows for a direct comparison of $\sigma_{ij}$ with the Navier-Stokes solver, inaccessible in the lattice-based LBM due to the staircase approximation of boundaries\cite{Stahl2010,Matyka2013}.

\subsection{Convergence of permeability and drag coefficient}\label{ssec:convergence}

We next check the convergence of the permeability $k/d^2$ (Eq.~\eqref{eq:k_definition}) and the drag coefficient $K$ with the space discretization refinement for various sphere radii. The drag coefficient is the non-dimensionalized drag force:
\begin{equation}\label{eq:K_definition}
    K = \frac{F_{H,0}}{6\pi r \varphi q \mu}.
\end{equation}
We use the minimal distance between nodes $h_\text{min}\in \{0.02, 0.014, 0.01, 0.007, 0.005\}$.

\begin{figure}[!ht]
\centering
\includegraphics[width=.95\linewidth]{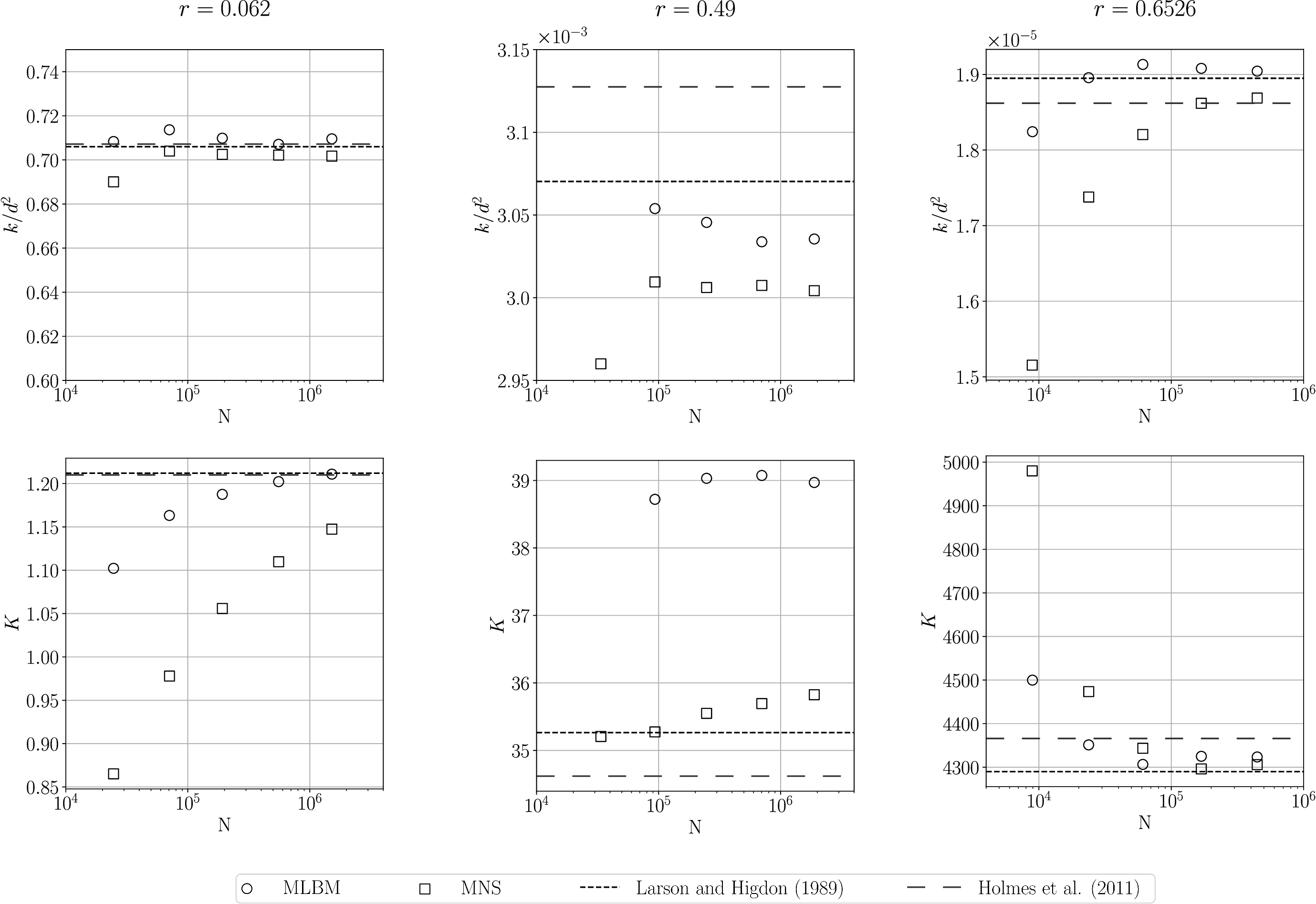}
\caption{The convergence of the dimensionless permeability $k/d^2$ and drag coefficient $K$ obtained with MLBM and MNS in 3D case for sphere radii $r=0.062$ (\textit{left}), $r=0.49$ (\textit{middle}) and $r=0.6526$ (\textit{right}). The horizontal dashed lines are the values obtained by Larson and Higdon \cite{Larson1989} and Holmes et al. \cite{Holmes2011} (interpolated for $r=0.49$).}
\label{fig:CONVERGENCE}
\end{figure}

The results are presented in Fig.~\ref{fig:CONVERGENCE} along with literature data ~\cite{Larson1989,Holmes2011}. For $r=0.49$ (near the touching limit) we interpolate the reference values of $K$ from the above works. To calculate the reference $k/d^2$ we use the relation:
\begin{equation}
    k/d^2 = \frac{1}{K}\left(\frac{d}{6\pi r}\right).
\end{equation}

For $r=0.062$, in the considered range of discretization parameters and for both tested methods, the values of permeability $k/d^2$ comply with the reference values within about $1\%$ relative difference. The drag coefficient $K$ converges towards the reference values in a stable manner. The significant discrepancy in convergence of $K$ for the MNS is most likely a result of pressure oscillations seen in Fig.~\ref{fig:pressure_cp} and further discussed in Sec.~\ref{ssec:pointwise_compariosn}.

For $r=0.49$, both methods converge to values of the permeability slightly off the reference values ($<\!3\%$ relative difference). The results of the drag force coefficient show a much higher discrepancy for MLBM (about $10\%$ compared to the reference). We note that for MLBM the $h_\text{min}=0.02$ ($N\approx 3\cdot 10^4$) case went unstable. To obtain stable solutions, the streaming distance needed to be changed to $\delta x=0.005$ for $h_\text{min}=0.014$ and the stencil size to $N_L=26$ for $h_\text{min}=0.01$ and to $N_L=35$ for $h_\text{min}=0.007$. The possible causes of the instability are further investigated in Sec.~\ref{ssec:svf_sweep}.

When large spheres are considered, permeability, as well as the drag coefficient obtained with both methods, converge to values close to those from the references. However, bigger differences from the reference values for coarse discretizations are visible in MNS. For the three considered cases we provide a quantitative analysis of the convergence in Appendix~\ref{sec:convergence_quantitative}.

Concerning those results, one can assess the efficiency advantage of the use of space discretization refinement as follows. The work of Holmes and others\cite{Holmes2011} reports the internodal distance approximately equal to $\tilde{h}=4.1 \cdot 10^{-3}$ for the porosity $\varphi=0.1$ ($r=0.6526$). Our results show that to achieve similar values of $k/d^2$ and $K$ for this case we need to have $h_\text{min}=7 \cdot 10^{-3}$, which gives $N\approx 1.7 \cdot 10^5$ nodes. Assuming the least dense, simple cubic packing of points in Holmes and others' work one can estimate the lower bound for the number of nodes in their simulation as about an order of magnitude higher:
\begin{equation}
    \tilde{N} = \frac{\varphi d^3}{\tilde{h}^3} = \frac{0.1 \cdot 1}{68.9 \cdot 10^{-9}} = 1.45 \cdot 10^6.
\end{equation}
Further in the text, we show that the number of nodes in the meshless solutions can be further decreased with no significant rise in the error of $K$.

\subsection{Permeability and drag coefficient for various porosities}\label{ssec:svf_sweep}

\begin{figure}[!ht]
\centering
\includegraphics[width=.75\linewidth]{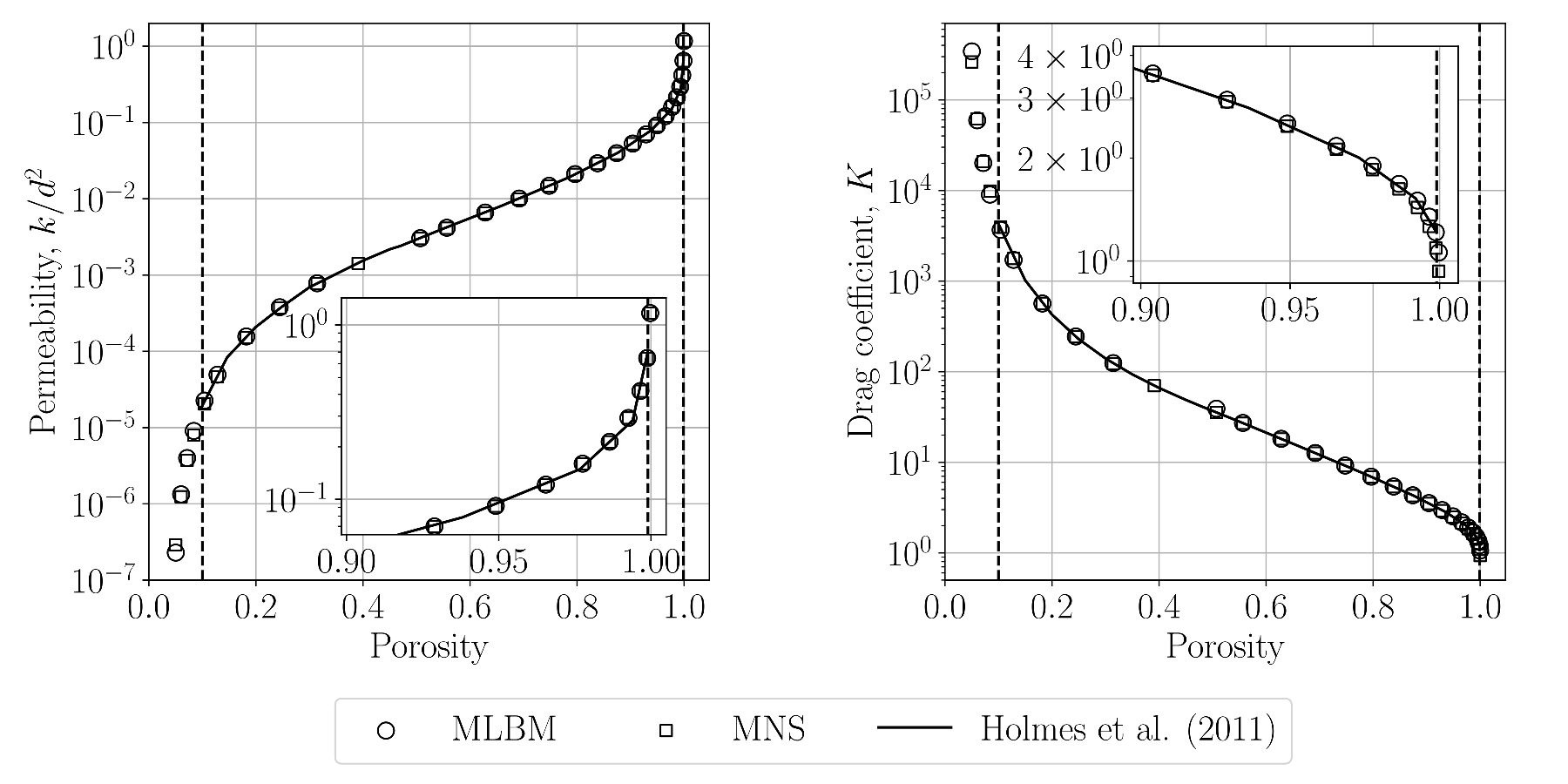}
\caption{The values of the dimensionless permeability $k/d^2$ and drag coefficient $K$ obtained with MLBM and MNS as a function of porosity. The solid line denotes the values obtained by Holmes et al. \cite{Holmes2011} in the range from $\varphi=0.1$ to $\varphi=0.999$ (indicated by vertical dashed lines).}
\label{fig:3d_vsHolmes_lbm}
\end{figure}

Next, we calculate the permeability and drag coefficient in a wide range of solid volume fractions. We choose the range $r \in [0.04,0.69]$. Fig.~\ref{fig:3d_vsHolmes_lbm} shows the values of $k/d^2$ and $K$ obtained for $h_\text{min}=0.01$ compared to the results of Holmes and others\cite{Holmes2011}. An excellent match between the two methods and the reference values is seen. This can be referred to the results of He and others \cite{He2002} where steady-state LBM and ACM solutions of the velocity and the pressure field are compliant with each other. MLBM and MNS are capable of extending the analysis range beyond those investigated by the previous authors (denoted by red dashed lines) without any changes to the refinement function, Eq.~\eqref{eq:refinement_function}. The meshless discretization with a proper refinement of the the narrow pore throats could open the possibilities to consider even lower porosities. The lack of MLBM data at $\varphi=0.4$ ($r=0.528$) was caused by the divergence of the simulation. We note that for $r = 0.49$ we used the stencil size $N_L=26$ for MLBM simulations.

\begin{figure}[!ht]
\centering
\includegraphics[width=.75\linewidth]{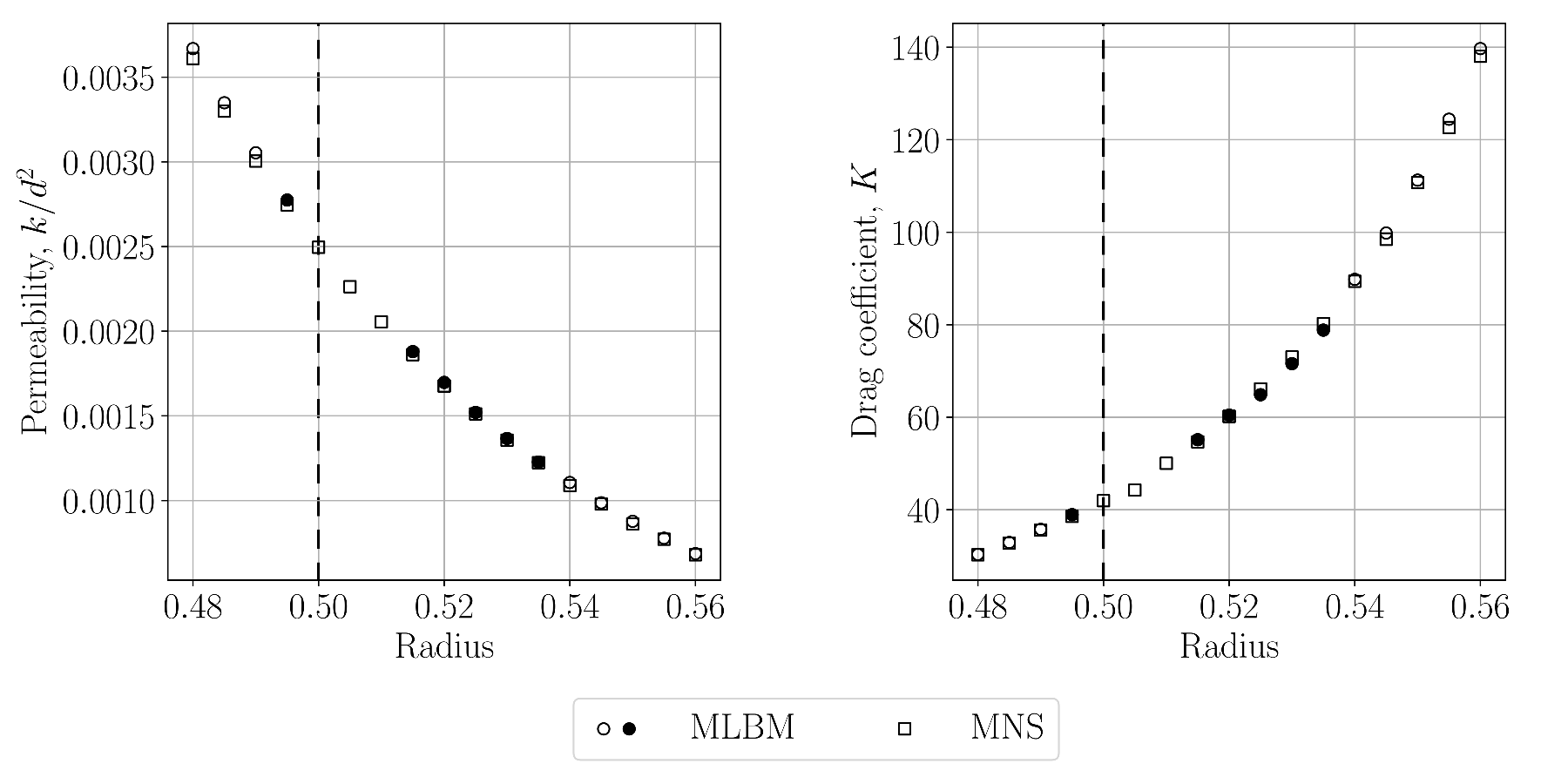}
\caption{The values of the permeability (\textit{left}) and the drag coefficient (\textit{right}) in the near-touching limit cases. The filled symbols denote the MLBM cases where the bounceback criterion or the interpolation parameters needed to be changed in order to obtain stable simulations. The dashed line indicates the touching limit.}
\label{fig:TOUCHING_LIMIT}
\end{figure}

To further investigate this issue we show the permeability and the drag coefficient values obtained for the near-touching limit cases ($r\approx0.5$) in Fig.~\ref{fig:TOUCHING_LIMIT}. For MLBM we use the discretizations with $h_\text{min}=0.007$ and the streaming distance length $\delta x=0.0035$. Away from the $r=0.5$ the values obtained with both methods are compliant with one another. For $r\rightarrow0.5$ MNS faces no problems with the solution stability, nor does it exhibit irregular behavior of the coefficients' values. On the other hand, MLBM simulations go unstable for the cases $r=0.5,0.505$ and $0.510$. To obtain a stable solution for radii close to the problematic range (i.e. $0.495 \le r \le 0.535$) we need to increase the stencil size (maximal considered value -- $N_L=45$) and change the rule for determining which populations of the boundary nodes are assumed unknown during the streaming. We no longer use the local normal vectors at the boundary nodes, rather we explicitly check which Lagrangian nodes lie inside the solid. Those altered MLBM setups are denoted with filled symbols in Fig.~\ref{fig:TOUCHING_LIMIT}. We attribute the difference in the behavior of the two methods in the touching limit case to the implementation of the no-slip boundary condition. In MNS, the Dirichlet boundary $\bsym{v}=\bsym{0}$ needs no approximation or information about the orientation of the local normal on the boundary, thus it introduces no error to the velocity field. MLBM faces growing approximation errors as the spheres get closer to each other and in the narrow volumes adjacent to the obstacle's intersections for $r>0.5$ which might produce unbalanced stencils due to visibility criterion~\cite{jacquemin_unified_2021}. Those can be further amplified by too few discretization points across the narrow fluid volumes compared to the minimal number of nodes used in the literature in e.g. Poiseuille flow\cite{He1997,Krueger2016}.

\subsection{Node placement sensitivity}
\begin{figure}[!ht]
\centering
\includegraphics[width=\linewidth]{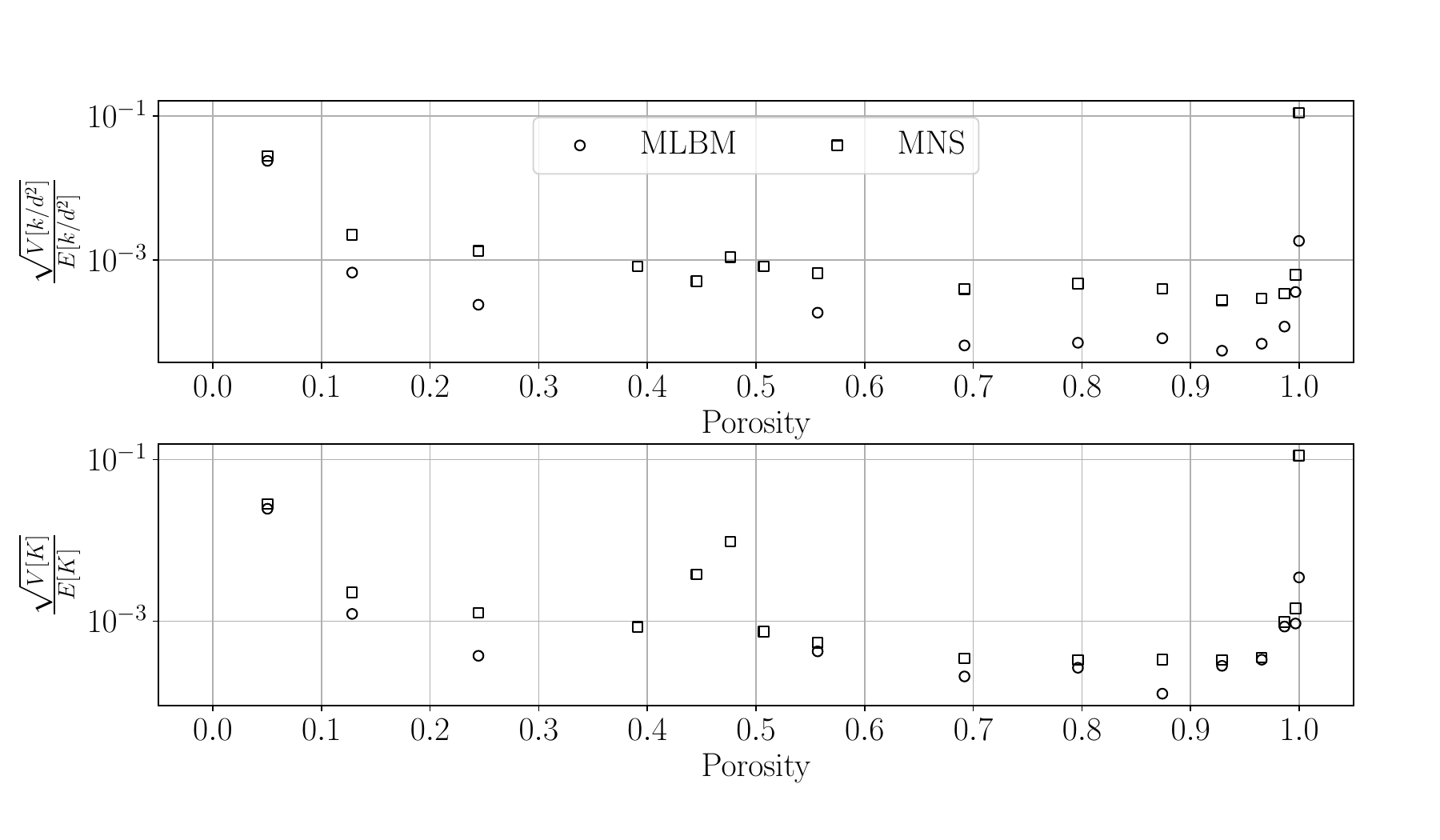}
\caption{The standard deviations of permeability (\textit{top}) and drag coefficient (\textit{bottom}) values normalized by the mean values of those parameters obtained on five realizations of discretizations for each radius of spheres. The realizations differed only in the value of the initial seed input to the discretization algorithm and were the same for both solvers.}
\label{fig:spread}
\end{figure}

Meshless methods inherently exhibit variability, as there are virtually infinite possible node layouts for a given target internodal distance~\cite{Strnisa2025}. This stems from differences in the number and placement of seed nodes, the number of expansion candidates, and the use of different randomization (random seeds) during candidate generation (see Section~\ref{ssec:meshless}). In this section, we analyze the effect of this variability on the results. In particular, we consider the systems of porosities in the same range as in Section~\ref{ssec:svf_sweep} discretized using $h_\text{min}=0.01$. For each sphere radius, we generate 5 discretizations differing from one another only in the value of the initial seed, the same for both solvers. We execute simulations with the stencil size $N_L=25$ on such a set of point clouds and for each porosity we calculate the standard deviation of the obtained values of $k/d^2$ and $K$ normalized by the estimated mean, see Fig.~\ref{fig:spread}. First of all, one sees that MLBM could not obtain converged results for all 5 realizations in several radii closest to the touching limit ($\varphi\sim0.48$), thus those values are not shown in the Figure. On the other hand, MNS successfully reached the steady state in all considered cases. This corresponds to the results shown in Figs.~\ref{fig:CONVERGENCE}--\ref{fig:TOUCHING_LIMIT}, where -- at least with the default approximation setup -- the problems with stability in MLBM were also evident. For the radii for which the converged results were obtained with each method, the values of the relative standard deviation are confined within the range $[5 \cdot 10^{-5};0.1]$. The three limiting cases -- extremely low and high porosity, as well as the touching limit -- give significantly higher values of the relative standard deviation of the coefficients. This is most likely caused by the small size of the discretized geometry details compared to the boundary internodal distance $h_\text{min}$, meaning that the individual placements play a much larger role. The surface of the smallest considered sphere with $r=0.04$ is discretized with $\sim175$ nodes, less that $0.01\%$ of all nodes, and the narrow throat in the case with the largest considered $r=0.69$ sphere is spanned by $\sim 4$ nodes.This directly points to the fact that those areas are bottlenecks for the precision of the simulations and local refinement there should be performed to obtain less varying results. In the intermediate porosities, MLBM solutions seem to be less dependent on the nodes placement than those of MNS.

\subsection{Influence of the refinement parameters and the stencil size}\label{ssec:refinement}

\begin{figure}[!ht]
\centering
\includegraphics[width=.75\linewidth]
{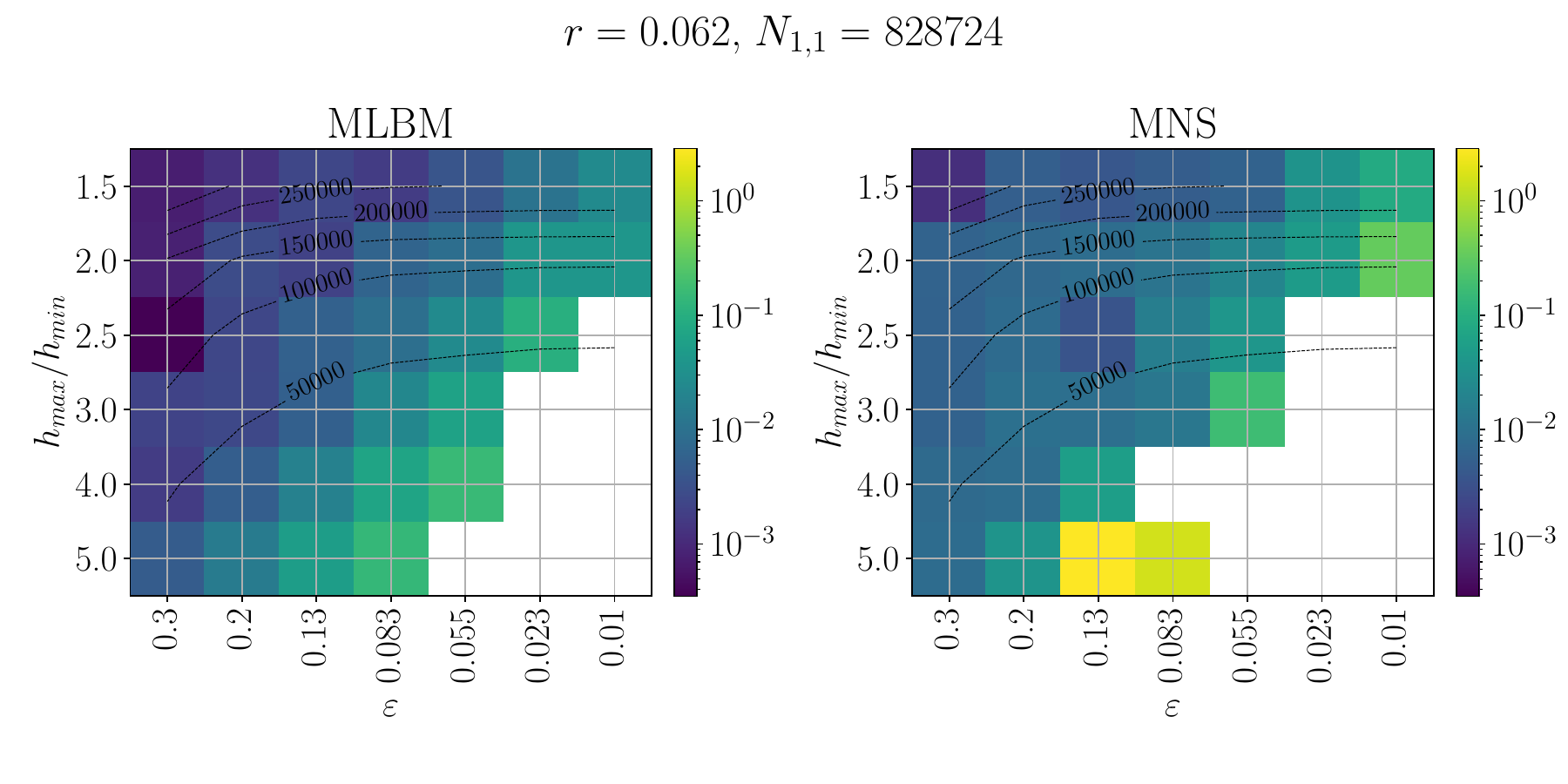}
\includegraphics[width=.75\linewidth]{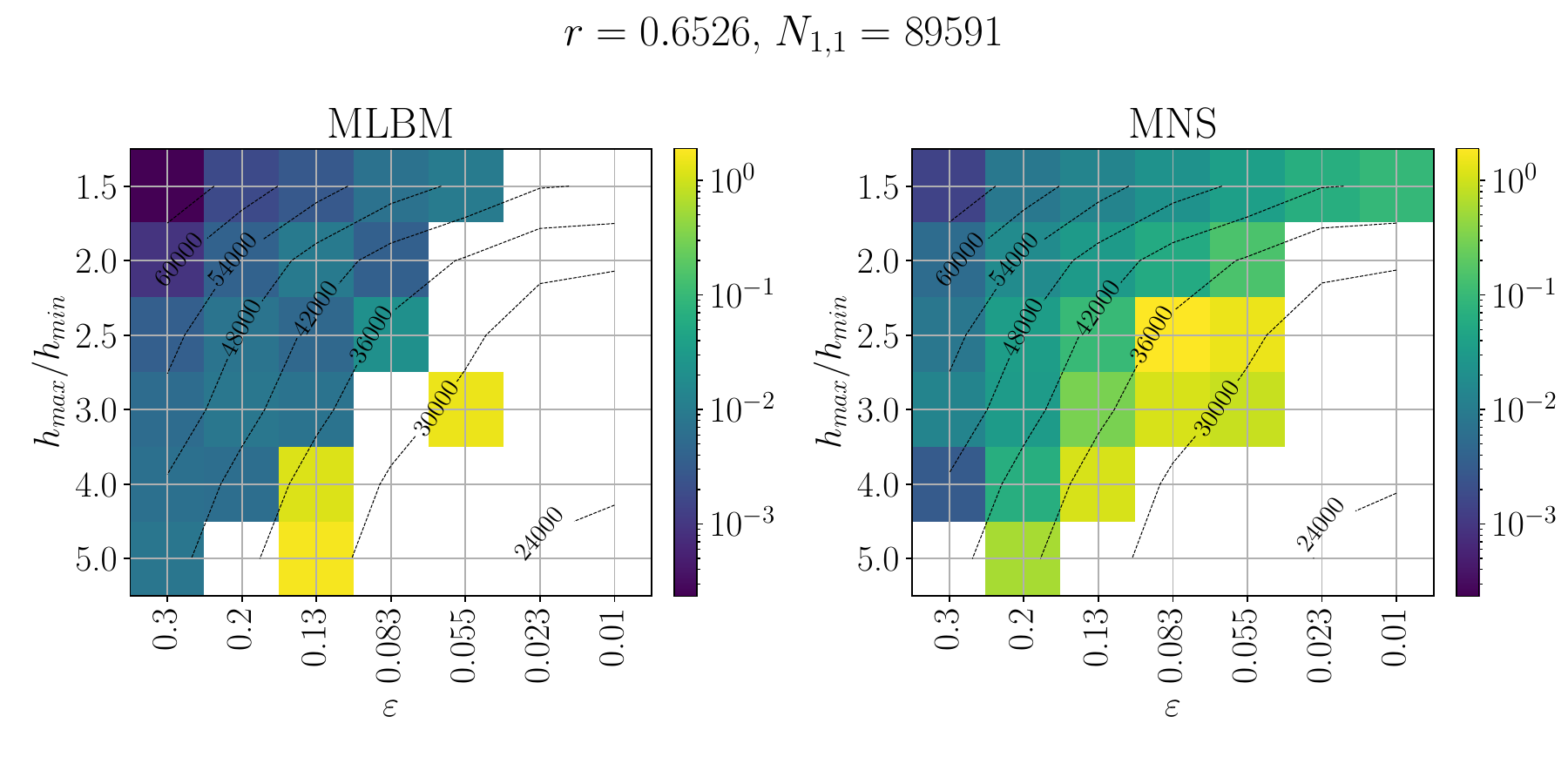}
\caption{Relative error of the drag coefficient for MLBM and MNS solution (Eq.~\eqref{eq:refinement_error}) for various refinement function parameters. $N_{1,1}$ is the number of nodes in the non-refined discretization. Dashed lines with labels are isocontours of the number of Eulerian nodes $N$.}
\label{fig:refinement_error}
\end{figure}

The advantage of meshless discretizations lies in the flexible positioning of nodes and easy manipulation of the approximation accuracy via the number of stencil members. First, we investigate the stability of the simulations and the relative errors of the drag coefficient $K$ obtained with various combinations of $\varepsilon$ and $h_\text{max}/h_\text{min}$ ratio (Eq.~\eqref{eq:refinement_function}) for small and large spheres. In general, lower values of $\varepsilon$ and higher values of $h_\text{max}/h_\text{min}$ correspond to more aggressive refinement from the bulk towards the sphere's surface (Fig.~\ref{fig:hFunctions}). The relative error, defined as:
\begin{equation}\label{eq:refinement_error}
    E\left(h_\text{max}/h_\text{min},\varepsilon\right) = \frac{|K\left(h_\text{max}/h_\text{min},\varepsilon\right) - K_\text{ref}|}{K_\text{ref}},
\end{equation}
is shown in Fig.~\ref{fig:refinement_error}. $K_\text{ref}\equiv K(1,1)$ is the value of $K$ for the constant nodes density in space. For all simulations, we used $h_\text{min}=0.01$.  No error value is plotted for setups when the simulations diverged or when $E>2$. On top of the errors map we plot isolines of several chosen numbers of nodes in the domain $N$. For comparison, in Fig.~\ref{fig:refinement_error} we also report the number of nodes for the non-refined case, $N_{1,1}$.

For the considered refinement parameters, both methods show similar regions of stable solutions. This suggests that this is by large the meshless approximation stability/accuracy which determines the stability for both methods. The values of error $E$ approximately follow the isocontours of the number of nodes with a noticeable drop of the error for large $\varepsilon$ at a constant $h_\text{max}/h_\text{min}$ ratio in the case of $r=0.062$. This is most probably due to a too small volume of fine discretization near the sphere to properly handle gradients in the relatively thick boundary layer. In this case, MLBM exhibits an extended area of stability compared to MNS. For the convergent cases of $r=0.6526$, the error values are comparable between the methods. For $r=0.6526$, MNS exhibits a larger set of stable solutions for low $h_\text{max}/h_\text{min}$ and $\varepsilon$ than MLBM, with higher values of errors in the range of medium values of those parameters. A comparison with the number of nodes in the non-refined case ($N_{1,1}$) reveals that the local refinement brings great advantages in the case of high-porosity sample -- approximately $15$-fold reduction of the node count comes with the cost of less than $1\%$ error for $h_\text{max}/h_\text{min}=4$, $\varepsilon=0.3$ case. Also, in the high-porosity system, increasing the shape parameter $\varepsilon$ (making the refinement function less peaked) up to the value of $\approx 0.2$ lowers the relative error $E$ with little change in the number of nodes. It suggests that increasing the volume near the walls where the discretization is refined is more beneficial than increasing the nodes density on the wall, which points to the proper resolution of gradients near the obstacle while the solution in the bulk varies slowly in space. In the low-porosity system, the number of nodes depends equally strong on both refinement measures, thus the increase in the computational burden is rather inevitable when the precision is to be improved.

\begin{figure}[!ht]
\centering
\includegraphics[width=.75\linewidth]{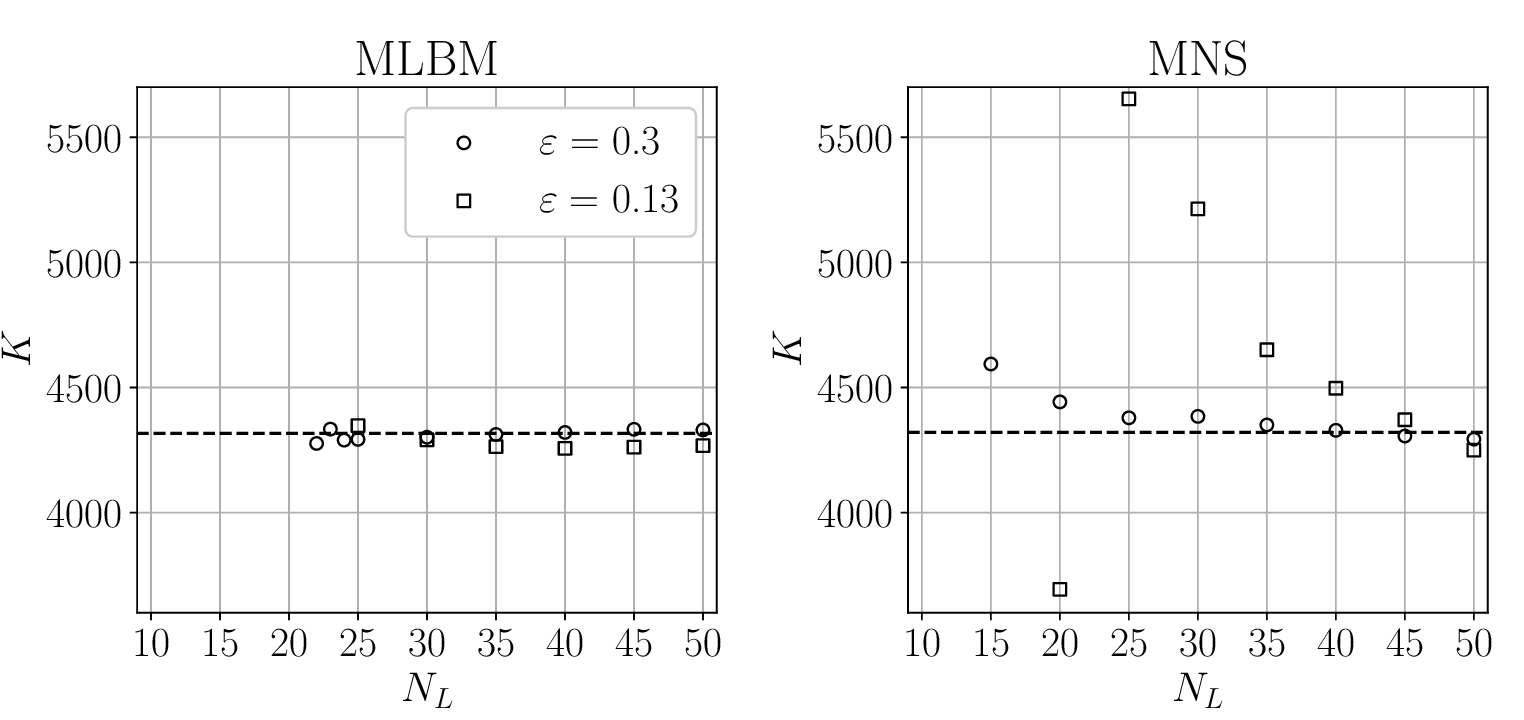}
\caption{The values of the drag coefficient $K$ for $r=0.6526$, $h_\text{max}/h_\text{min}=3$ case versus the stencil size $N_L$ for various shape parameters $\varepsilon$ for both solvers. The dashed lines are the reference $K_\text{ref}\equiv K(1,1)$ values.}
\label{fig:stencil_size_analysis}
\end{figure}

Fig.~\ref{fig:stencil_size_analysis} shows the obtained values of the drag coefficient for various stencil sizes $N_L$ and refinement function shape parameters $\varepsilon$. In the case of MLBM, the value of $K$ shows no significant dependence on the stencil size, regardless of the refinement aggressiveness. MNS can provide stable solutions with smaller stencil sizes but exhibits a stronger dependence of $K$ on $N_L$. This can be explained by the fact that smaller $\varepsilon$, i.e. large gradients in node density, result in unbalanced stencils due to them being constructed from $N_L$ closest nodes irrespective of direction. Unbalanced stencils are not as problematic for the RBF-FD method in value interpolation used by MLBM but can cause significant problems for 1st and 2nd-order derivative approximation used in MNS.

\subsection{Pointwise comparison of the meshless solvers}\label{ssec:pointwise_compariosn}

\begin{figure}[!ht]
\centering
\includegraphics[width=0.45\linewidth]{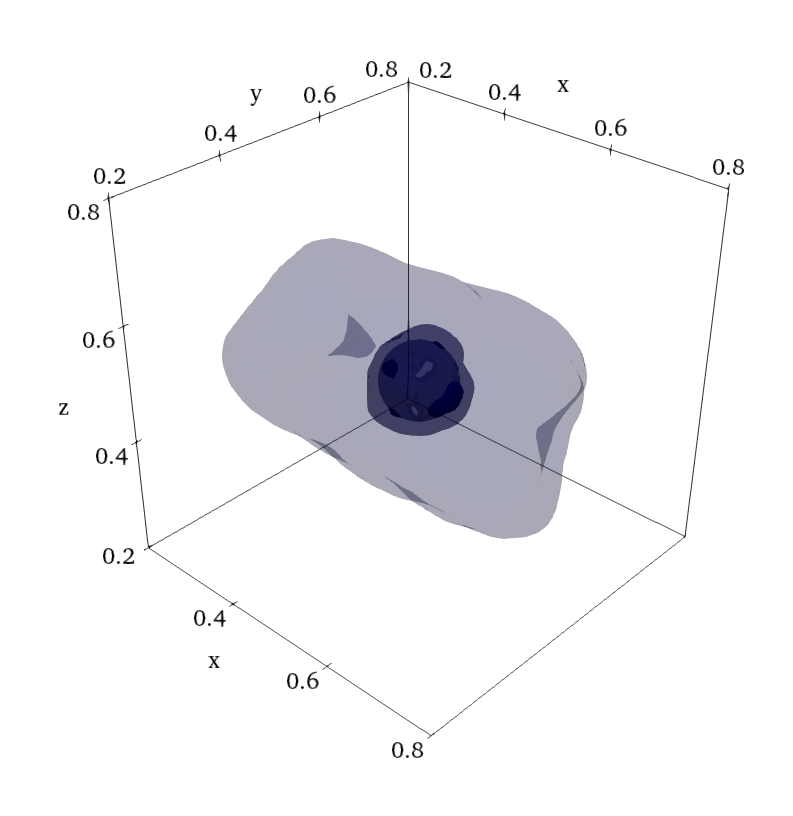}
\hspace{0.5cm}
\includegraphics[width=0.45\linewidth]{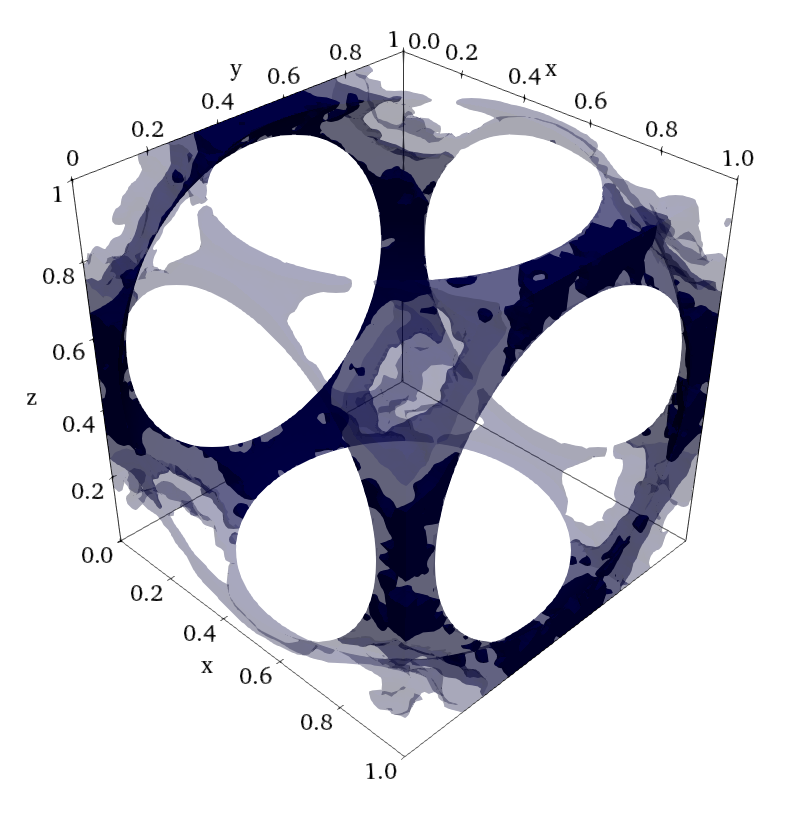}
\caption{Relative pointwise velocity difference between the two solvers $\delta (\bsym{x})$ for $r=0.062$ (\textit{left}) and $r=0.6526$ (\textit{right}). The isocontours correspond to the values: $0.1,0.05,0.01$ for $r=0.062$ and $1,0.25,0.05$ for $r=0.6526$, from the darkest to the brightest. Note the sphere surface is not rendered but the isosurfaces are trimmed by it.}
\label{fig:pointwise_comparison}
\end{figure}

\begin{figure}[!ht]
\centering
\includegraphics[height=0.35\linewidth]{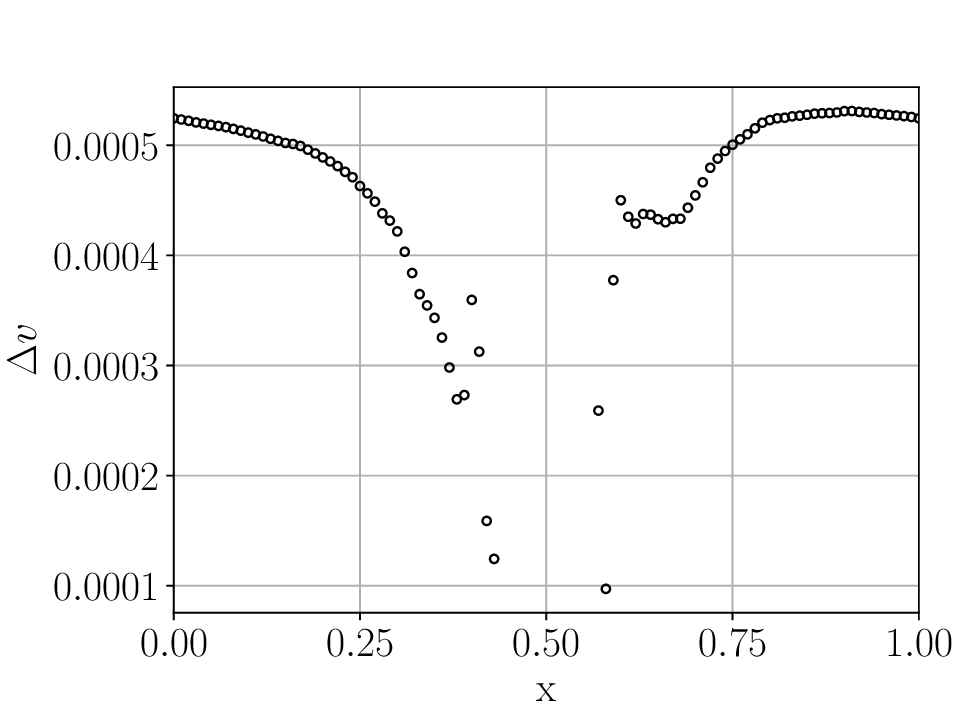}
\hspace{0.5cm}
\includegraphics[height=0.35\linewidth]{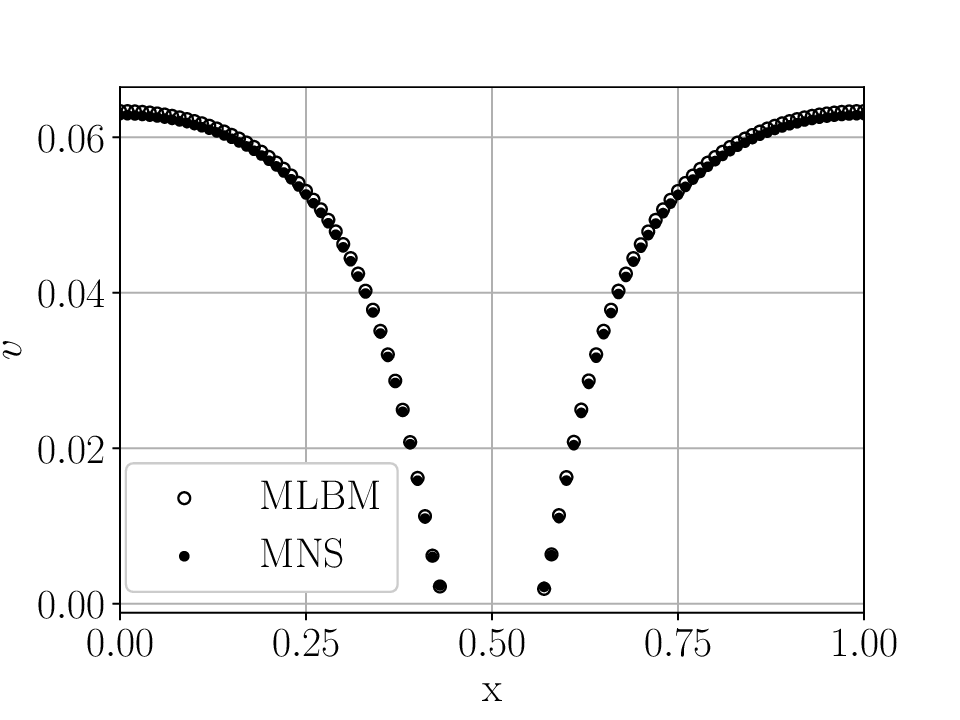}
\caption{The absolute value of the difference of the velocity magnitude between MLBM and MNS $\Delta v$ (\textit{left}) and the values of the velocity magnitude for each method (\textit{right}) along the line $y,z=0.5$ for the $r=0.062$ case.}
\label{fig:pointwise_comparison_line}
\end{figure}

We next investigate the pointwise relative difference between the velocity fields obtained with MLBM and MNS. A single simulation with $r=0.062$ and $r=0.6526$ was run with each solver using exactly the same space discretization. We define the pointwise difference of the velocity field as:
\begin{equation}\label{eq:pointwise_error}
    \delta(\bsym{x}) = \frac{|\bsym{v}_\text{MLBM}-\bsym{v}_\text{MNS}|}{|\bsym{v}_\text{MNS}|} = \frac{\Delta v}{|\bsym{v}_\text{MNS}|}
\end{equation}
where the subscript denotes the method used to obtain the velocity value and $|\cdot|$ is the Euclidean vector norm. Fig.~\ref{fig:pointwise_comparison} shows isosurfaces of $\delta(\bsym{x})$ for each sphere radius. For $r=0.062$ the lowest difference isosurface corresponds to the boundary layer around the sphere, being stretched in the $x$-direction. The larger values of $\delta(\bsym{x})$ towards the surface are caused by the normalization by near-zero velocity values (compare with the right subplot of Fig.~\ref{fig:pointwise_comparison_line} showing that the absolute value of the difference falls down towards the surface) with the maxima occurring in the same places as the maxima of the tangent force of MNS solution in Fig.~\ref{fig:pressure_cp}. More drag in the MNS solution may explain the slightly lower $x$-velocity visible in the right plot of Fig.~\ref{fig:pointwise_comparison_line}. In the case of $r=0.6526$ the largest differences occur in the concave parts of the domain, near the spheres' intersections, where vortices are expected to emerge in the velocity field, even for a very low Reynolds number \cite{Agnaou2017}. The fact that the main point-wise differences shown in Fig.~\ref{fig:pointwise_comparison} and Fig.~\ref{fig:pointwise_comparison_line} occur in areas with stagnant flow indicates that it may be caused by known ACM problems with pressure oscillations in the low Reynolds regime\cite{clausen2013EDAC}. This is supported by the oscillations in the surface pressure field shown in Fig.~\ref{fig:pressure_cp}, where the fluctuations in pressure isocountours appear noticeably larger in areas where the tangent force is lower, indicating a stagnant flow.

\subsection{Complexity and timings analysis}

\noindent\textit{Algorithmical complexity}

\begin{figure}[!ht]
\centering
\includegraphics[width=.49\linewidth]
{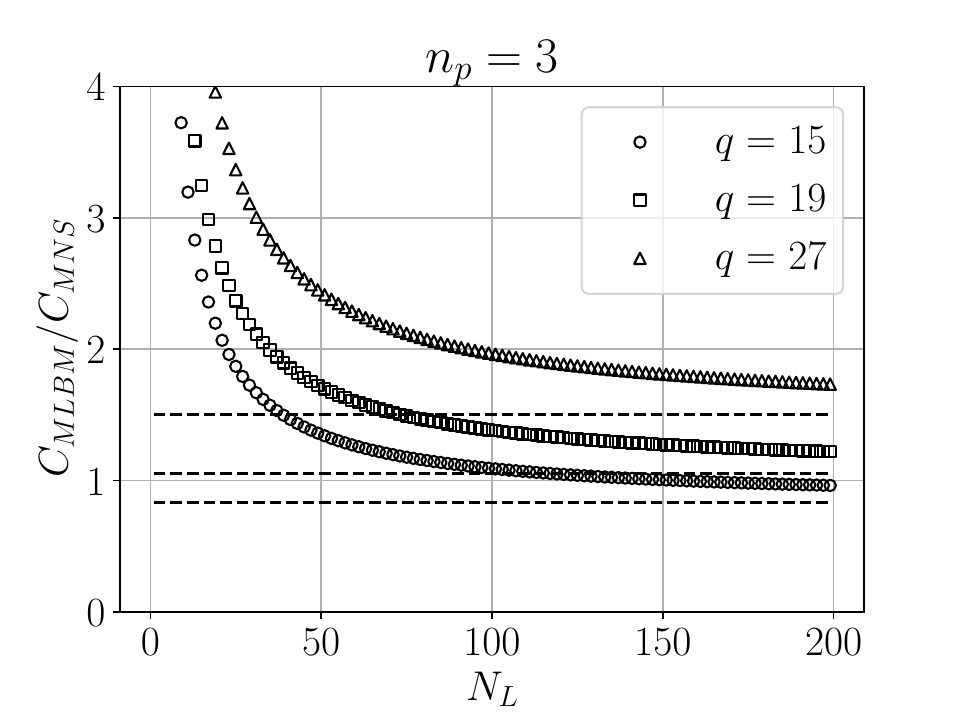}
\includegraphics[width=.49\linewidth]{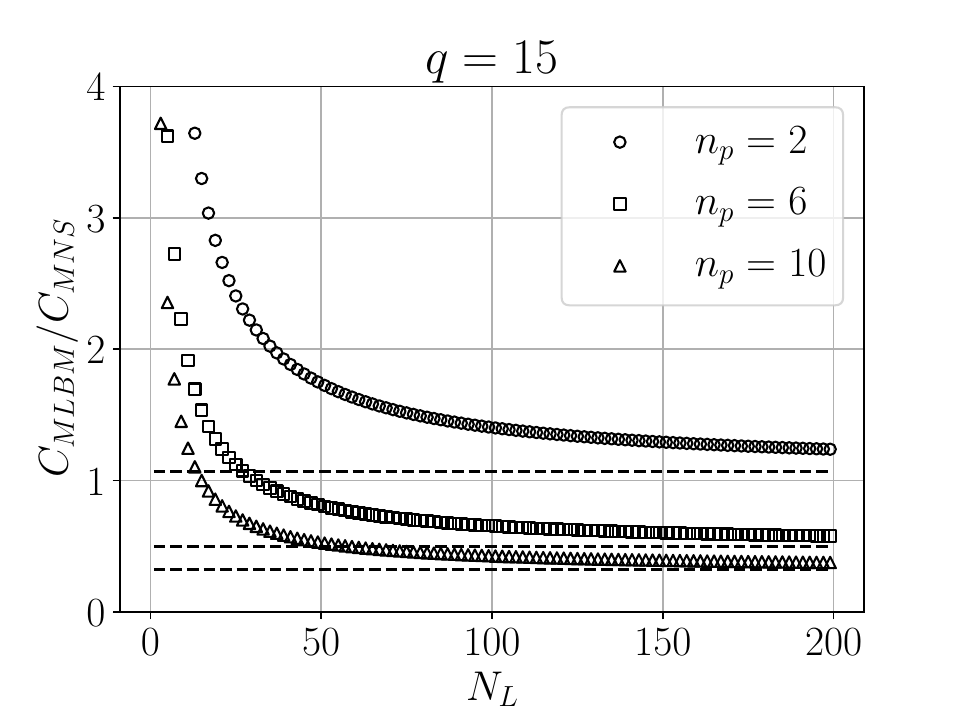}
\caption{The ratio of MLBM complexity to MNS complexity defined by Eqs.~\eqref{eq:complexity_mlbm} and \eqref{eq:complexity_mns} for three velocity discretizations used in MLBM (\textit{left}) and for a range of pressure correction step numbers $n_p$ (\textit{right}). The horizontal dashed lines denote the asymptotes of the ratio for $N_L \rightarrow +\infty$.}
\label{fig:complexity_ratio}
\end{figure}

The cost of a single timestep of the described MLBM algorithm with the TRT collision kernel can be expressed as:
\begin{equation}\label{eq:complexity_mlbm}
    \begin{split}
        C_\text{MLBM} = N \left(q(\underbrace{56}_\text{I} + \underbrace{2N_L-1}_\text{II}) + \underbrace{q-1 \> + \> 6q}_\text{III}\right) = N(q(62+2N_L)-1)\\
    \end{split}
\end{equation}
where terms denoted by I, II, and III correspond to the collision, the interpolation, and the calculation of the macroscopic variables according to Eq.~\eqref{eq:macro_var}, respectively. In the case of the present MNS implementation, it is:
\\
\begin{equation}\label{eq:complexity_mns}
    C_\text{MNS} = N\left(\underbrace{6(2N_L-1)+4}_\text{I} + \underbrace{n_p (4(2N_L-1) + 4)}_\text{II}\right) = N (N_L (8 n_p + 12) - 2)
\end{equation}
where the terms denoted by I and II correspond to the update of the velocity field and the pressure correction steps, respectively. It is immediately seen that the number of the stencil members $N_L$ contributes significantly to both values. In MLBM its contribution is multiplied by the number of discrete velocities $q$, while in MNS by terms corresponding to the number of macroscopic fields and pressure correction steps $n_p$. In MLBM, $q$ multiplies also a large factor of $62$ related to the calculation of the equilibrium distributions. In fact, the asymptotic complexity of each method is $\mathcal{O}(NN_Lq)$ for MLBM (in line with the findings of e.g. Musavi and Ashrafizaadeh \cite{musavi2015meshless}) and $\mathcal{O}(NN_Ln_p)$ for MNS.  However, for the asymptotic regimes to be reached, extremely large values of $N_L$, $q$ and $n_p$ would need to be used. It is therefore more useful to consider the values of complexities for some certain, smaller values of those parameters. Fig.~\ref{fig:complexity_ratio} shows the ratio of the two complexities for a constant number of pressure correction steps $n_p$ and discrete velocities $q$. The complexity of both methods depends on the choice of the free parameters. For small stencils, the overhead from the calculation of equilibrium in MLBM makes it more complex than MNS.

It is important to notice that the size of the monomial subset of the interpolation basis does not influence the approximation step's complexity directly. It does, however, impose limits on the minimal number of nodes in interpolation supports, as noted in Section~\ref{ssec:meshless}. The approximation step makes use of the weights vectors, Eq.~\eqref{eq:augmentedSystem}, which need to be calculated only once, in the preprocessing step. The computational cost of this is of the order of $\mathcal{O}(N(N_L+m)^3)$ which has to be multiplied by the number of populations in MLBM or the number of approximated operators in MNS.
\\

\noindent \textit{Timings}

\begin{figure}[!ht]
\centering
\includegraphics[width=.75\linewidth]{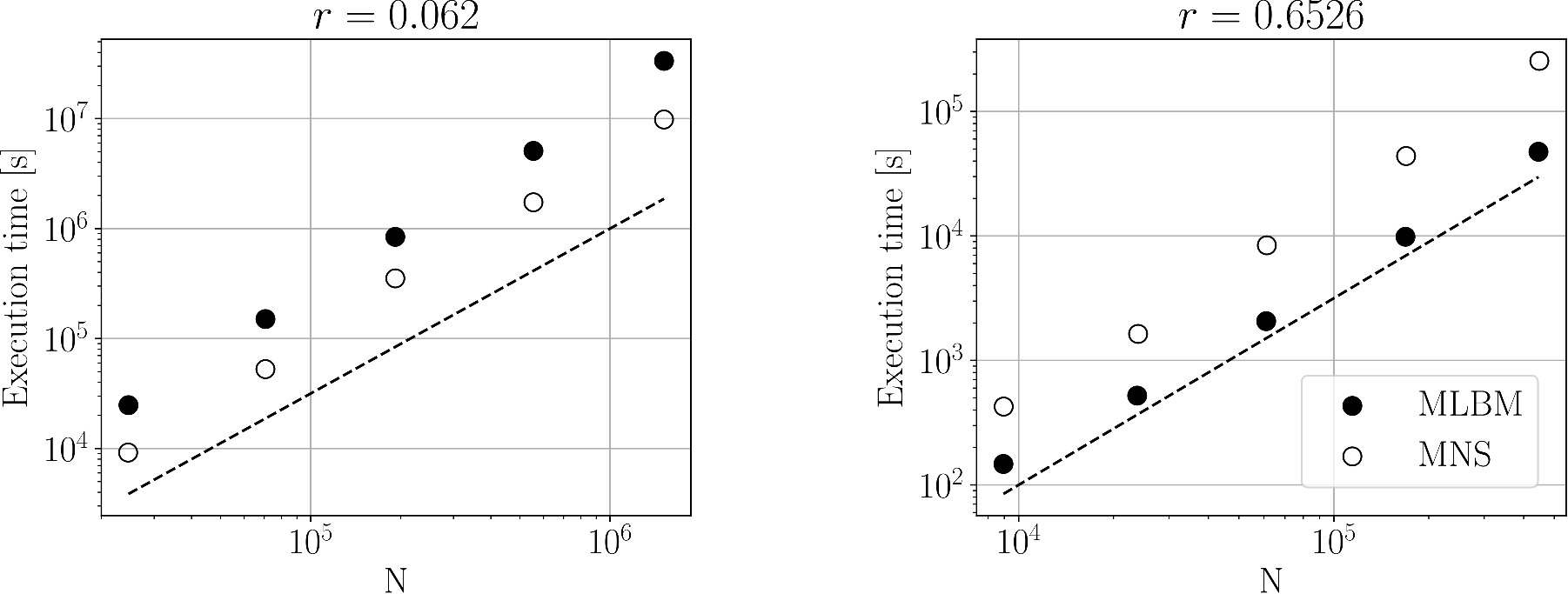}
\caption{Execution times in seconds of $r=0.062$ (\textit{left}) and $r=0.6526$ (\textit{right}) case for single-threaded runs versus the number of nodes in the domain $N$. Open symbols denote meshless Navier-Stokes timings and filled symbols denote MLBM timings. The stopping criterion was the same as in the remaining tests. In both methods, the execution time is inversely proportional to the timestep length $\delta t$. In both methods the diffusive scaling of the timestep length was used ($\delta t \propto h(\bsym{x})^2$ in MNS and $\delta t \propto \delta x^2 \propto h(\bsym{x})^2$ in MLBM) and as $\delta x \propto N^{1/3}$ one obtains that the execution time should scale as $N^{1.5}$. The dashed lines denote the 1.5-order slope. The execution times for $r=0.062$ were extrapolated based on the physical time at which the simulation reached the stopping criterion and the execution time for a fixed number of iterations.}
\label{fig:timings}
\end{figure}

To provide a more practical quantification of the performance of the presented methods we measure the time needed to achieve a converged solution (defined as in the previous tests) with each of them. Fig.~\ref{fig:timings} shows the algorithms' execution times for the case of $r=0.062$ and $r=0.6526$ versus the number of nodes in the discretization $N$. We perform the analysis for single-threaded runs on 2x Intel E5520 (2.27 GHz) with 12GB of RAM machine. We observing that the scaling is of the order $1.5$ for both methods and compliant with the theoretical predictions. In the large sphere case, the MNS needs approximately $3$-folds more time to complete the calculations compared to the MLBM. We note, however, that the presented execution time implicitly contains the amount of the physical time needed for the system to reach steady-state. MLBM needed about $3.57$s and $0.04$s to reach the stopping criterion for $r=0.062$ and $r=0.6526$, respectively, at $h_\text{min}=0.007$. Those times for MNS are $3.58$s and $0.39$s. In the case of $r=0.6526$, the difference of $0.35$s in the physical time is what accounts for the larger execution times of MNS in Fig.~\ref{fig:timings}. In case $r=0.062$, where the physical time of the simulations is comparable between the methods, the estimated execution times were about three times larger for MLBM than for MNS. The issues of pressure oscillations mentioned in Sec.\ref{ssec:pointwise_compariosn} may explain the difference in the physical time required to reach a steady state in the $r=0.6526$ case with large areas of stagnant flow.

\section{Conclusions}
%%%%%%%% 2024-04-02 VERSION %%%%%%%%
The paper compares two meshless CFD solvers for the simulation of fluid flow through porous media, the meshless Lattice-Boltzmann method with two relaxation-time collision terms and a direct Navier-Stokes solver under the artificial compressibility limit, focusing on a periodic 3D porous medium modeled as a cubic array of spheres. Both methods utilize point clouds of the same point density in space, provided by an iterative, advancing-front dimension-independent algorithm that allows variable internodal distances (needed for $h$-refinement) and Radial Basis Function (RBF) approximation for the field and operator approximation. In terms of methods, the paper discusses the meshless discretization and approximation techniques as well as the specifics of the implementation of the considered CFD methods.

Through a convergence analysis of the permeability and drag coefficient analysis across various porosities, we demonstrate the effectiveness of both methods in accurately predicting the flow characteristics. Both methods are convergent and provide results that agree with reference data from the literature. This has encouraged us to extend the range of solvable problems beyond the available benchmark data to lower porosities.

Furthermore, we investigate the sensitivity of the drag coefficient to the refinement parameters, with both methods showing a similar stability range. We demonstrate the advantages of meshless refined discretization by significantly reducing the number of nodes with minimal increase in error.

When analyzing the effect of the stencil size, we find that the meshless Lattice-Boltzmann method achieves good results even with a relatively small stencil size compared to the direct solver. This is to some extent to be expected, since the Lattice-Boltzmann method uses the RBF approximation only for field approximation, while the direct solver approximates all differential operators involved.

The paper also delves into the complexity and timing analyses of both solvers. We find the complexity of both methods linearly dependent on the stencil size and the space discretization size with coefficients related to the equilibrium VDF form (LBM) and the number of pressure steps (MNS). Although the MNS is seemingly less complex in terms of the sheer number of operations, it requires a larger stencil size for stability reasons. Moreover, the actual execution times strongly depend on the number of iterations required to reach the steady state, which is different for both methods for different porosities. The message here is that both methods are in the same order of magnitude in terms of computational complexity.

\section{Data availability}
The full raw datasets and codes used to generate them and analyzed during the current study are available in two public repositories: 
https://zenodo.org/records/14979637 and 
https://zenodo.org/records/14979739.

\section{Acknowledgments}
The authors acknowledge the financial support from the Slovenian Research and Innovation Agency (ARIS) research core funding No. P2-0095 and the Young Researcher programme PR-10468.
Funded by National Science Centre, Poland under the OPUS call in the Weave programme 2021/43/I/ST3/00228.
This research was funded in whole or in part by National Science Centre (2021/43/I/ST3/00228). For the purpose of Open Access,
the author has applied a CC-BY public copyright licence to any Author Accepted Manuscript (AAM) version arising from this submission.

\section*{Author contributions statement}
D.S. developed the meshless LBM method, M.R. developed a meshless NS solver, and G.K. and M.M. formulated the research problem. G.K. and M.M. organized funding and equipment to perform computations. All authors analyzed the results. All authors reviewed the manuscript. 

\appendix

\section{MLBM and MNS algorithms comparison}\label{app:algorithms_comparison}

\begin{minipage}{0.46\textwidth}
    \centering MLBM algorithm
    \flushleft
    % \hline
    \begin{algorithmic}[1]
        \While{the stopping criterion is not met}
        \State \texttt{// COLLIDE}
        \For{all Eulerian points $\bsym{x}_i$}
        \For{each lattice velocity $\bsym{c}_k$}
        \State Calculate $f^\text{eq}_k$ (Eq.~\eqref{eq:feq});
        \State Calculate $f^\text{post}_k$ (Eq.~\eqref{eq:LBM_collision});
        \EndFor
        \EndFor
        \State \texttt{// STREAM}
        \For{all Eulerian points $\bsym{x}_i$}
        \For{each Lagrangian point $\bsym{x}_i+\delta\bsym{x}_k$}
        \State Interpolate $f^\text{post}_{k'}$ to $\bsym{x}_i+\delta\bsym{x}_k$ (Eq.~\eqref{eq:operatorApprox});
        \State Overwrite $f_{k'}(\bsym{x}_i)$ with $f^\text{post}_{k'}(\bsym{x}_i+\delta\bsym{x}_k)$ (Eq.~\eqref{eq:LBM_streaming});
        \EndFor
        \State Update $\rho_{lb}$, $\bsym{v}_{lb}$ (Eq.~\eqref{eq:macro_var});
        \EndFor
        \EndWhile
    \end{algorithmic}
    \vfill
\end{minipage}
\hfill
\begin{minipage}{0.46\textwidth}
    \centering MNS algorithm
    \flushleft
    % \hline
    \begin{algorithmic}[1]
        \While{the stopping criterion is not met}
        \State \texttt{// UPDATE VELOCITY}
        \For{all computational points $\bsym{x}_i$}
        \State Calculate intermediate velocity  $\vec{v}'$ (Eq.~\eqref{eq:MNS_intermediate});
        \EndFor
        \State \texttt{// PRESSURE-VELOCITY COUPLING}
        \For{a predetermined number of iterations}
        \For{all computational points $\bsym{x}_i$}
        \State Update pressure (Eq.~\eqref{eq:MNS_pressure});
        \EndFor
        \For{all computational points $\bsym{x}_i$}
        \State Update velocity with the new pressure gradient Eq.~\eqref{eq:MNS_velocity};
        \State Update $C$ if required (Eq.~\eqref{eq:MNS_C});
        \EndFor
        \EndFor
        \EndWhile
        \State
    \end{algorithmic}
    \vfill
\end{minipage}

\section{Quantitative analysis of convergence}\label{sec:convergence_quantitative}

To provide a quantitative description of the convergence we calculate  the grid convergence index (GCI) for permeability and drag coefficients, $\text{GCI}_{k/d^2}$ and $\text{GCI}_K$ respectively. GCI is calculated for the values of permeability or drag coefficient obtained on the finer discretization in each pair of subsequently refined grids. For example, if the pair of the discretizations $h_\text{min}=0.014$ and $h_\text{min}=0.01$ is considered, we calculate the GCI for the discretization $h_\text{min}=0.01$ and so on. For each such pair, GCI is calculated as~\cite{}
\begin{equation}\label{eq:gci_convergence_study}
    \text{GCI}_X = \frac{F_s |\epsilon_X|}{r^{p_X}-1}
\end{equation}
where $r>1$ is the ratio between the coarser and the finer $h_\text{min}$ within the pair. $p_X$ is the apparent order of convergence of the quantity $X$ obtained from a linear regression fit to the $( \> \log{(h_\text{min})},\log{(e_X(h_\text{min}))} \> )$ data points where $e_X$ is the relative error for a quantity $X$ calculated as
\begin{equation}\label{eq:relative_error_convergence_study}
    e_X(h_\text{min}) = \frac{|X(h_\text{min}) - X(h_\text{min}=0.005)|}{X(h_\text{min}=0.005)},
    \quad
    X \in \{k/d^2,K\}
\end{equation}
and $F_s=1.3$ is the safety factor. The relative error for quantity $X$ in each pair, $\epsilon_X$, is calculated as
\begin{equation}\label{eq:pairwise_relative_error_convergence_study}
    \epsilon_X = \frac{|X_\text{coarser} - X_\text{finer}|}{X_\text{finer}}
\end{equation}
where subscripts \textit{coarser} and \textit{finer} denote the value obtained on the coarser and the finer of the two discretizations in the pair. We present the values of GCI in Tables~\ref{tab:gci_r0-062}--\ref{tab:gci_r0-6526}. Due to the lack of data for MLBM at $r=0.49$ for the coarsest discretization, we do not include the $h_\text{min}=0.02$ points in the calculation of GCI for both methods in general.

The GCI study confirms the observations from Fig.~\ref{fig:CONVERGENCE}. If the highest values on the finest pair of discretizations for each method are considered, these are those for drag coefficient in $r=0.062$ and $r=0.49$ cases for MLBM ($7.2 \cdot 10^{-3}$ and $6.8 \cdot 10^{-3}$, respectively) and for drag coefficient at $r=0.062$ and permeability at $r=0.49$ for MNS ($3.9 \cdot 10^{-2}$ and $5.0 \cdot 10^{-3}$, respectively). This suggests that to obtain grid-independent results, one would need much denser discretizations in those setups.

\begin{table}[!ht]
    \centering
    \begin{tabular}{ccc}
        % --- MLBM ---
            \begin{tabular}{lrcccc}
                $h$ & \multicolumn{1}{c}{$N$} & $\text{GCI}_{k/d^2}$ & $\text{GCI}_K$ \\
                \hline
                0.014 & 70641      & \multicolumn{1}{c}{--} & \multicolumn{1}{c}{--} \\
                0.01  & 191802     &      7.9e-03 &      2.1e-02 \\
                0.007 & 555077     &      5.1e-03 &      1.1e-02 \\
                0.005 & 1515847    &      5.4e-03 &      7.2e-03 \\
            \end{tabular}
        & \hspace{1cm} &
        % --- MNS ---
            \begin{tabular}{lrcccc}
                $h$ & \multicolumn{1}{c}{$N$} & $\text{GCI}_{k/d^2}$ & $\text{GCI}_K$ \\
                \hline
                0.014 & 70785    & \multicolumn{1}{c}{--} & \multicolumn{1}{c}{--} \\
                0.01  & 191896   &      2.3e-03 &      8.9e-02 \\
                0.007 & 555333   &      4.8e-04 &      5.4e-02 \\
                0.005 & 1516144 &      7.0e-04 &      3.9e-02 \\
            \end{tabular}\\
        \end{tabular}
    \caption{$r=0.062$: Relative errors and GCI for permeability and drag coefficient. \textit{Left:} MLBM results, \textit{right:} MNS results. The MLBM permeability was for each simulation fitted in time with a function of the form $a+b\exp{(-ct)}$ and the asymptote $a$ was taken to be the actual value of the permeability.}
    \label{tab:gci_r0-062}
\end{table}

\begin{table}[!ht]
    \centering
    \begin{tabular}{ccc}
        % --- MLBM ---
            \begin{tabular}{lrcccc}
                $h$ & \multicolumn{1}{c}{$N$} & $\text{GCI}_{k/d^2}$ & $\text{GCI}_K$ \\
                \hline
                0.014 & 93348      & \multicolumn{1}{c}{--} & \multicolumn{1}{c}{--} \\
                0.01  & 247355     &      1.6e-03 &      2.0e-02 \\
                0.007 & 701396     &      2.0e-03 &      2.5e-03 \\
                0.005 & 1887779    &      3.2e-04 &      6.8e-03 \\
            \end{tabular}
        & \hspace{1cm} &
        % --- MNS ---
            \begin{tabular}{lrcccc}
                $h$ & \multicolumn{1}{c}{$N$} & $\text{GCI}_{k/d^2}$ & $\text{GCI}_K$ \\
                \hline
                0.014 & 93348    & \multicolumn{1}{c}{--} & \multicolumn{1}{c}{--} \\
                0.01  & 247355   &      5.4e-03 &      1.0e-02 \\
                0.007 & 701396   &      1.9e-03 &      4.8e-03 \\
                0.005 & 1887779 &      5.0e-03 &      4.7e-03 \\
            \end{tabular}\\
        \end{tabular}
    \caption{$r=0.49$: Relative errors and GCI for permeability and drag coefficient. \textit{Left:} MLBM results, \textit{right:} MNS results.}
    \label{tab:gci_r0-5}
\end{table}

\begin{table}[!ht]
    \centering
    \begin{tabular}{ccc}
        % --- MLBM ---
            \begin{tabular}{lrcccc}
                $h$ & \multicolumn{1}{c}{$N$} & $\text{GCI}_{k/d^2}$ & $\text{GCI}_K$ \\
                \hline
                0.014 & 23792.0    & \multicolumn{1}{c}{--} & \multicolumn{1}{c}{--} \\
                0.01  & 60950.0    &      1.9e-02 &      1.3e-02 \\
                0.007 & 168569.0   &      8.1e-03 &      4.4e-03 \\
                0.005 & 445457.0   &      1.5e-03 &      1.0e-03 \\
            \end{tabular}
        & \hspace{1cm} &
        % --- MNS ---
            \begin{tabular}{lrcccc}
                $h$ & \multicolumn{1}{c}{$N$} & $\text{GCI}_{k/d^2}$ & $\text{GCI}_K$ \\
                \hline
                0.014 & 23767      & \multicolumn{1}{c}{--} & \multicolumn{1}{c}{--} \\
                0.01  & 60988      &      2.3e-02 &      4.2e-03 \\
                0.007 & 168704     &      6.3e-03 &      1.6e-03 \\
                0.005 & 445717     &      4.8e-03 &      1.4e-04 \\
            \end{tabular}\\
        \end{tabular}
    \caption{$r=0.6526$: Relative errors and GCI for permeability and drag coefficient. \textit{Left:} MLBM results, \textit{right:} MNS results.}
    \label{tab:gci_r0-6526}
\end{table}

\section{List of symbols used in the text}\label{ssec:list_of_symbols}

\begin{table}[!ht]
    \centering
    \begin{tabular}{cc}
        \begin{tabular}{l p{.35\textwidth}}
            \hline
            \multicolumn{2}{c}{Greek letters} \\
            \hline
    
            $\beta$ & compressibility parameter \\
            $\delta$ & pointwise difference of the velocity field between the two solvers \\
            $\delta t$ & timestep length \\
            $\delta x$ & MLBM streaming distance \\
            $\delta_i$ & local scaling factor for radial functions \\
            $\delta_{ij}$ & Kronecker delta \\
            $|\Delta|_{k/d^2}$ & relative change of permeability in time \\
            $\varepsilon$ & refinement function shape parameter \\
            $\epsilon_X$ & relative difference between the values of quantity $X$ obtained on two subsequently refined discretizations \\
            $\theta_s$ & polar angle on the obstacle's surface \\
            $\lambda$ & monomial weights vector \\
            $\mu$ & dynamic viscosity \\
            $\nu$, $\nu_{lb}$ & kinematic viscosity, kinematic viscosity in LB units \\
            $\rho$, $\rho_{lb}$ & density, density in LB units \\
            $\rho_\text{ref}$ & reference density \\
            $\sigma_{ij}$ & stress tensor \\
            $\tau$ & non-dimensional relaxation time \\
            $\tilde{\phi}$ & normalized signed distance function \\
            $\phi_s$ & azimuthal angle on the obstacle's surface \\
            $\phi_\text{sdf}$ & signed distance function \\
            $\varphi$ & porosity \\
            $\Phi$ & radial function \\
            $\omega_k$ & $k$-th lattice weight \\
            $\Omega$ & volume of the fluid in the domain \\
            
            \hline
            \multicolumn{2}{c}{Latin letters} \\
            \hline
    
            $A$ & radial functions part of the approximation matrix or area of the solid walls \\
            $dA$, $\Delta A$ & infinitesimal and finite surface element \\
            $b$, $c$ & radial functions and monomials part of the right-hand side vector of the approximation problem, respectively \\
            $c_s$ & lattice speed of sound \\
            $C$ & artificial speed of sound \\
            $C_\text{MLBM}$ & computational complexity of MLBM \\
            $C_\text{MNS}$ & computational complexity of MNS \\
            $d$ & computational domain's side length \\
            $D$ & dimensionality of the problem \\
            $e_k$, $e_{k'}$ & $k$-th discrete streaming vector, vector opposite to $e_k$ \\
        \end{tabular} &
        \begin{tabular}{l p{.35\textwidth}}
            $E$ & relative error of drag coefficient in the refinement parameters study \\
            $f_k$, $f_k^\text{eq}$, $f_k^\text{post}$ & $k$-th discrete velocity distribution function (VDF), equilibrium VDF, post-collisional VDF \\
            $F_{H,i}$ & $i$-th component of the hydrodynamic force acting on the obstacles \\
            $F_k$ & $k$-th discrete body force term \\
            $F_s$ & safety factor in grid convergence index calculation \\
            $g$, $g_{lb}$ & body force, body force in LB units \\
            $GCI_X$ & grid convergence index of quantity $X$ \\
            $h$ & local internodal distance \\
            $h_\text{min}$, $h_\text{max}$ & minimal and maximal internodal distance \\
            $\tilde{h}$ & approximate internodal distance \\
            $I_b$ & set of boundary nodes indices \\
            $k$ & order of the polyharmonic spline radial function \\
            $k/d^2$ & dimensionless permeability \\
            $K$, $K_\text{ref}$ & drag coefficient, drag coefficient obtained on non-refined discretization \\
            $\mathcal{L}$ & linear differential operator \\
            $m$ & number of monomials augmenting the meshless approximation matrix \\
            $n$ & boundary normal direction \\
            $\hat{n}_i$ & local boundary normal vector \\
            $n_p$ & number of pressure correction steps \\
            $N$, $N_{1,1}$ & number of points in the refined and non-refined discretization \\
            $\tilde{N}$ & approximate number of nodes in the referenced works \\
            $N_b$ & number of the solid boundary nodes \\
            $N_L$ & stencil size in meshless approximation \\
            $N_p$ & number of monomials augmenting the approximation matrix \\
            $p$ & pressure \\
            $p_l$ & $l$-th monomial augmenting the approximation matrix \\
            $p_X$ & order of convergence of quantity $X$ \\
            $P$ & monomial part of the augmented approximation matrix \\
            $P_i$ & $i$-th cross-section for calculating the mean $x$-component of velocity \\
        \end{tabular} \\
    \end{tabular}
    \caption*{List of symbols used in the text (1/2)}
\end{table}

\begin{table}[!ht]
    \centering
    \begin{tabular}{cc}
        \begin{tabular}{l p{.35\textwidth}}
            \hline
            \multicolumn{2}{c}{Latin letters (cont.)} \\
            \hline

            $q$ & number of discrete velocities in the Lattice Boltzmann Method model or mean $x$-component of velocity \\
            $r$ & radial function argument or obstacle radius \\
            $S_i$, $S_i(j)$ & stencil of the $i$-th node, $j$-th member of $S_i$ \\
            $t$ & time \\
            $\Delta t$ & time interval for calculating the relative change of permeability in time \\
            $u$ & approximated function \\
            
        \end{tabular} &
        \begin{tabular}{l p{.35\textwidth}}

            $v$, $v_{lb}$ & macroscopic velocity, macroscopic velocity in LB units \\
            $v_{ref}$ & reference macroscopic velocity \\
            $\Delta v$ & absolute value of the difference of the velocity magnitude between MLBM and MNS \\
            $w_{i,j}$ & approximation weight of the $j$-th node in the $i$-stencil \\
            $x$, $x_i$ & space coordinate \\
            $x_d$ & de-queued node position in the node placing algorithm \\
            
        \end{tabular} \\
    \end{tabular}
    \caption*{List of symbols used in the text (2/2)}
\end{table}

\bibliography{sample}

\end{document}